\documentclass[11pt,a4paper]{article}

\pdfoutput=1
\usepackage{jheppub}
\usepackage{simplewick}
\usepackage{thumbpdf}
\usepackage{amsmath}
\usepackage{amssymb}
\usepackage{array}
\usepackage{cancel}
\usepackage[hang,small]{caption2}
\usepackage{subfigure}

\newenvironment{rcases}
  {\left.\begin{aligned}}
  {\end{aligned}\right\rbrace}
  
\newcommand{\be}{\begin{equation}}
\newcommand{\ee}{\end{equation}}

\title{Correlators of chiral primaries and 1/8 BPS Wilson loops from perturbation theory}
\author{Marisa Bonini,}
\author{Luca Griguolo}  
\author{and Michelangelo Preti}
\affiliation{Dipartimento di Fisica, Universit\`a di Parma and INFN Gruppo Collegato di Parma, \\Viale G. P. Usberti 7/A, 43100 Parma}
\emailAdd{marisa.bonini@fis.unipr.it} 
\emailAdd{luca.griguolo@fis.unipr.it} 
\emailAdd{michelangelo.preti@fis.unipr.it} 
\abstract{We study at perturbative level the correlation functions of a general class of 
1/8 BPS Wilson loops and chiral primaries in ${\cal N}=4$ Super Yang-Mills theory. 
The contours and the location of operator insertions share a sphere $S^2$ embedded 
into space-time and the system preserves at least two supercharges. We perform 
explicit two-loop computations, for some particular but still rather general 
configuration, that confirm the elegant results expected from localization procedure. 
We find notably full consistency with the multi-matrix model averages, obtained from 
2D Yang-Mills theory on the sphere, when interacting diagrams do not cancel and 
contribute non-trivially to the final answer.}

\begin{document}
\maketitle


\section{Introduction and results}
In recent years localization has been proven to be one of the most
powerful tool in obtaining non perturbative results in quantum
supersymmetric gauge theories
\cite{Nekrasov:2002qd,Nekrasov:2003rj,Pestun:2007rz,Kapustin:2009kz}. The
key point is that supersymmetry algebras can be often deformed to
accommodate background curvature on compact spaces and the resulting
partition functions can be computed via a particular saddle-point
procedure, known as the supersymmetric localization technique. Thanks to
this procedure, an impressive number of new exact results have been
derived for supersymmetric theories in different dimensions, mainly when
formulated on spheres or products thereof.
The technique is enough flexible to compute also correlation functions of
local operators and expectation values of non local observables, such as
Wilson loops \cite{Pestun:2007rz,Kapustin:2009kz} and 't Hooft loops
\cite{Gomis:2011pf,Drukker:2012sr}. Actually the exact expression for
circular 1/2 BPS Wilson loops in ${\cal N}=4$ Super Yang-Mills theory was
conjectured \cite{Erickson:2000af,Drukker:2000rr} long before its concrete derivation
through supersymmetric localization.
This procedure in turn generalizes to a large class of ${\cal N}=2$
theories, where Wilson loops can be also accurately studied
\cite{Passerini:2011fe,Russo:2012ay,Bigazzi:2013xia} through matrix model
techniques.

In  the case of ${\cal N}=4$, the 1/2-BPS circle can be generalized to Wilson
loops of arbitrary shapes with lower degree of supersymmetry
\cite{Drukker:2007qr} (for a complete classification see
\cite{Dymarsky:2009si,Cardinali:2012sy}). A particular family within this
construction is composed by arbitrary loops lying on a two-sphere $S^2$
embedded into the Euclidean spacetime. These operators are generically
1/8-BPS, and their quantum correlators seem to be reproduced exactly by a purely
perturbative calculation in bosonic 2D Yang-Mills
\cite{Drukker:2007qr,Bassetto:1998sr}. The original conjecture was
substantially proved\footnote{The Wilson loop operators localize on a 2D
gauge theory similar to the Hitchin/Higgs-Yang Mills system
\cite{Moore:1997dj,Gerasimov:2006zt,Gerasimov:2007ap}, that is
perturbatively equivalent to the usual two-dimensional Yang-Mills theory.}
using supersymmetric localization \cite{Pestun:2009nn}. The computation in
the two-dimensional theory can be exactly mapped to simple Gaussian
multi-matrix models \cite{Giombi:2009ms}, leading to an explicit
evaluation of the correlators. The relation to 2D YM has been thourougly
checked
\cite{Bassetto:2008yf,Young:2008ed,Bassetto:2009rt,Bassetto:2009ms} and
extended to the inclusion of 't Hooft loops \cite{Giombi:2009ek}. Quite
interestingly the localization of this family of Wilson loops has been
instrumental in deriving a non perturbative expression for the so-called
Bremsstrahlung function \cite{Correa:2012at,Fiol:2012sg}, a non-BPS
quantity governing the radiation emitted by an accellerated quark in the
small velocity limit. The final result has also been tested using
integrability \cite{Gromov:2012eu,Gromov:2013qga}, providing a beautiful
relation between calculations performed through localization and
integrability.

More generally localization should apply not only to the Wilson and 't
Hooft loops, but to a whole sector of operators that are annihilated by
shared supercharges. In particular it should concern a family of chiral
primary operators inserted on $S^2$ \cite{Giombi:2009ds}, leading to exact
results for their correlators also in presence of Wilson loops. The
correlation function of a local operator and a Wilson loop in this sector
was firstly computed in \cite{Giombi:2009ds}, supporting and extending the
original conjecture of \cite{Semenoff:2001xp} for the correlator of a 1/2-BPS Wilson loop
and a chiral primary (see also \cite{Semenoff:2006am} for the study of the
1/4 BPS case). The correspondence with the zero-instanton sector of
two-dimensional YM was checked at tree level, finding consistency with the localization result.
In a further development \cite{Giombi:2012ep} the investigation of the protected
sector was extended to the realm of three-point functions. A careful
computation of the correlator of two chiral primaries on $S^2$ with a Wilson
loop of arbitrary shape was performed there using a Gaussian three-matrix
model. Large $R$-charge and strong coupling
limits were also explored, in order to make contact with the string picture,
and interesting results have been obtained considering one "heavy" and one
"light" primary \cite{Zarembo:2010rr,Costa:2010rz}. These calculations
should be considered important for recent advances on
three-point functions study through AdS/CFT duality and integrability
\cite{Escobedo:2010xs}.

In this paper we take instead a more conservative point of view and study
the same correlation functions, considered in \cite{Giombi:2009ds},
through the conventional diagrammatic expansion. Of course the first aim
is to check the highly non-trivial reorganization of the perturbative
series encoded into the two-matrix model result: localization
automatically performs a number of divergences cancellation among
different diagrams and combines finite contributions into nice
expressions, written in terms of the geometry of the correlator. These
effects are by no means obvious, expecially when the position of the
operator and the shape of the contour are arbitrary. The appearing of a
gaussian  matrix model suggests that only the combinatorics of
perturbation theory should mind when bosonic propagators connecting points on
the circuit are constant, as for the 1/2 BPS circle in Feynman gauge
\cite{Erickson:2000af}. In this case the contributions of the
interactions, coming from internal loops and non-trivial vertices, should
cancel among themselves. The first situation that we examine reproduces
exactly this pattern: we consider a chiral operator inserted on the
north-pole of $S^2$ and a Wilson loop placed on a latitude. The bosonic
propagators are constant and we check explicitly the complete
cancellation of the interacting diagrams at two-loops: we use dimensional
regularization to tame the divergences appearing in the intermediate steps
of the calculation and some Mellin-Barnes technology, adapted to our
integration contours, to compute the relevant graphs. The resummation of
the perturbative exchanges is then easily performed, leading to the
expected result. A more involved situation arises when the operator is
inserted in a arbitrary point of one of the two emispheres. The structure of
the operator itself changes and the bosonic propagators suspended on the
loop are no more constant, complicating the actual computation: in
particular the terms involving three contour integrations cannot be
reduced completely to double-integrals, as in the previous case.
Moreover to evaluate the interacting diagrams we must resort to numerical 
integration. These diagrams interplay with the ladder ones to reproduces the 
matrix model result.
In our computation we consider chiral primaries of dimension
$J=2$: at two-loop this is not really a limitation, in fact one can extend the
perturbative evaluation to the general case with some combinatorial
effort\footnote{The basic combination at two-loop level always involve
two-legs diagram, so $J=2$ is the most general situation at this order.}.

It would be possible to extend the present computation to the case of two
chiral primaries and one Wilson loop, checking in this way the expression
derived in \cite{Giombi:2012ep}. More generally one could try to develop
an analogous supersymmetric system in three-dimensional ABJM theory
\cite{Aharony:2008ug}, where a family of 1/6 BPS Wilson loops living on
the two-sphere $S^2$ with the same properties of the 1/8 BPS operators
considered here has been recently derived \cite{Cardinali:2012ru} and
studied at quantum level \cite{Bianchi:2014laa}. Chiral primaries sharing
part of the supersymmetries should be constructed and, at least at
perturbative level, the correlation function could be studied. We leave
these projects for the future.

The structure of the paper is the following. In section 2 we recall the
relevant operators (Wilson loops an chiral primaries) and the matrix
models describing their correlation functions. In section 3 we outline the
computation of the correlation function between a chiral primary inserted
at the north-pole and a latitude Wilson loop: in particular we organize
the diagrams and write down the result for the building blocks that
cancel among themselves. In section 4 we consider the case of a chiral
primary in an arbitrary position. We show that exchange diagrams do not
reproduce the matrix model answer and present the contribution of the
interaction, organized in basic building blocks. We have various
appendices devoted to technical aspects of the computations presented
in the body of the paper. 

\section{The localization result and multi-matrix models from 2D Yang-Mills}
The Wilson loops that we consider in this paper are generically 1/8-BPS
operators and have been constructed in \cite{Drukker:2007qr}. They are
supported on arbitrary closed curves on a $S^2$ embedded into the
Euclidean four-dimensional space. The relevant two-sphere is defined in
Cartesian coordinates as

\be
x_4 =0, \,\,\,\,\, \,\,\,\,\sum^{3}_{i=1} x^2_i =R^2.
\ee

In the following we will take $R=1$\footnote{ The dependence from the
radius of the two-sphere can be easily reintroduced.}. To
obtain 1/8-BPS Wilson loops one should engineer a suitable coupling with
three of the six scalars $\Phi^i$, $i= 1, 2, 3$, of ${\cal N}=4$ SYM and
for any contour ${\cal C}$ the explicit form of the operator is given by

\be\label{WLDGRT}
W_{\cal R}[{\cal C}]=\frac{1}{{\rm dim}_{\cal R}} {\rm Tr}_{\cal R} P\exp
i\int_{\cal C} \Bigl [ A_i +i \epsilon_{ijk}\Phi^j x_k \Bigr ] dx^i,
\ee
where
${\rm dim}_{\cal R}$ denotes the dimension of the representation ${\cal
R}$. 
Because the four supercharges preserved by the loops do not depend on the circuit,
a system of Wilson loops on $S^2$ is 1/8-BPS.
Supersymmetry enhances for special shapes: the well-known 1/2-BPS circular
Wilson loop is obtained by taking ${\cal C}$ to be an equator of $S^2$.
Circles of arbitrary radius along latitudes of $S^2$ are 1/4-BPS and they
coincide with the 1/4-BPS Wilson loops of \cite{Drukker:2006ga}. 

This is not the end of the story: we can also insert an arbitrary number
of local operators on the same $S^2$ still preserving two supercharges.
The  local operators doing the job are the following
\be
\mathcal{O}_J(x)\sim {\rm \text{Tr}}\Bigl(x_k\Phi^k(x) +i\Phi^4(x)\Bigr)^J
\,\,\,\,\,\,\,\,\,\, x_k \in S^2\,,\,\,\,\,\, k=1,2,3.
\ee
They are of course ordinary chiral primaries, the orientation in the
scalar space being simply correlated with the position of the insertion on
$S^2$. The two-point function of these operators is position
independent, as can be easily shown from the direct definition, and, upon
choosing a suitable normalization, it holds
\be
\langle \mathcal{O}_J(x)\,\mathcal{O}_{J'}(x)\rangle = \delta_{J
J'}.
\ee
More generally all the $n$-point functions $\langle
\mathcal{O}_{J_1}(x_1)\mathcal{O}_{J_2}(x_2)...\mathcal{O}_{J_n}(x_n)\rangle$
are position independent \cite{{Drukker:2009sf}} and tree-level exact.
Any collection of these operators on $S^2$ also preserves four supercharges. In presence
of the Wilson loops (\ref{WLDGRT}), the system becomes invariant under two
supercharges \cite{Giombi:2009ds} and mixed correlation functions of
Wilson loops and local operators can depend non-trivially on the coupling
constant, as we will discuss in the next sections. The two preserved
supercharges can be combined \cite{Giombi:2009ds} to obtain the fermionic
charge used in the localization procedure of \cite{Pestun:2009nn}. Mixed
correlators of Wilson loops and local operators should therefore be
exactly computed by the perturbative sector two-dimensional Yang-Mills
theory on $S^2$ \cite{Bassetto:1998sr}, according to the proposal of
\cite{Drukker:2007qr}.

Two-dimensional Yang-Mills theory can be exactly solved on any Riemann
surface, both using lattice \cite{Migdal:1975zf} and localization
\cite{Witten:1991we} techniques and its zero-instanton sector is
described by certain Gaussian matrix models \cite{Giombi:2009ms}. The
relevant four-dimensional correlators can be eventually mapped to
\be
\label{multi}
\langle W_{{\cal R}_1}[{\cal C}_1]  W_{{\cal R}_2}[{\cal
C}_2]...\mathcal{O}_{J_1} (x_1)\mathcal{O}_{J_2} (x_2) ... \rangle
= \frac{1}{{\cal Z}} \int [dX][dY] {\rm \text{Tr}}_{{\cal R}_1}
e^{X_1} {\rm \text{Tr}}_{{\cal R}_2} e^{X_2}...{\rm \text{Tr}}Y_1^{J_1}{\rm
\text{Tr}}Y_2^{J_2}...e^{S[X,Y]},
\ee
where the matrix model action $S[X,Y]$ is a quadratic form in $X_i$,$Y_i$
with coefficients depending on the areas singled out by the Wilson loops
and the topology of the system. 
We remarks that localization, and consequently the matrix
model description, does not need the large $N$ limit.

Special cases of the multi-matrix model (\ref{multi}) have been studied
and checked in the past: the case of a single Wilson loop has been tested
at two-loop \cite{Bassetto:2008yf,Young:2008ed} and at strong coupling
\cite{Drukker:2007qr} for a non-trivial wedge configuration. Correlators
of two Wilson loops were also considered
\cite{Bassetto:2009rt,Bassetto:2009ms} and explicit computations at order
$g^6_{_{YM}}$ have confirmed the matrix model result. The generic $n$-point
function for local operators has been studied in \cite{Giombi:2009ds}
where also the mixed correlator between a Wilson loop and a local operator
has been computed and studied in different regimes. Three-point functions
have been instead carefully scrutinized in \cite{Giombi:2012ep},
expecially at strong coupling and in relation with string computations.
Here we concentrate our attention on the mixed two-point correlators:
\be
\label{due}
\langle W_{{\cal R}}[{\cal C}]  \mathcal{O}_{J} (x_1)\rangle
= \frac{1}{{\cal Z}} \int [dX][dY] {\rm \text{Tr}}_{{\cal R}} e^{X} {\rm
\text{Tr}}Y^{J}e^{-\frac{A^2}{2g_{_{YM}}}{\rm
Tr}\bigl(\frac{A_1}{A_2}Y^2-\frac{2i}{A_2}XY
             \bigr)}.
\ee
Here $A_{1,2}$ are the areas single out by the loop on $S^2$
with $A=A_1+A_2$ and the local operator is
inserted into $A_1$. In this paper we will also consider
operators normalized as ordinary chiral primaries with unit two-point
function
\be\label{oj}
\mathcal{O}_J(x)=
\left(\frac{2\pi}{\sqrt{\lambda}}\right)^J\frac{1}{\sqrt{J}} {\rm
\text{Tr}}\Bigl(x_k\Phi^k(x) +i\Phi^4(x)\Bigr)^J.
\ee
Taking the trace of the Wilson loop in the fundamental representation and
considering the large $N$ limit, the matrix integral (\ref{due}) can be
readily done obtaining
\be\label{matrmodel}
\langle W_{{\cal R}}[{\cal C}]  \mathcal{O}_{J}
(x_1)\rangle=\frac 1 N\frac{\sqrt{J}}{2^J}\left(\frac{A_2}{A_1}\right)^{\frac{J}{2}}I_J(\sqrt{\lambda'}),
\ee
where $\lambda'=4\lambda A_1A_2/A^2$. We will
check this expression at second order in perturbation theory. For loops of
arbitrary shape and arbitrary operator insertion this result cannot be
recovered simply by resumming the ladder exchanges. On the other hand, to
perform a concrete computation, we must limit ourselves to some particular
configuration, keeping enough generality to observe non-trivially emerging
of the matrix model answer. We will consider two cases: in the first one
the operator is inserted at the north pole and the loop is placed at an
arbitrary latitude. Propagators are constant and interactions should cancel.
We then consider a second configuration, where the operator is inserted at
an arbitrary point and the loop is wrapped at the equator: here, as we
will see, interactions are expected to contribute non-trivially to the
final result.

\section{Perturbative computations I: latitude Wilson loop with an operator insertion at the north-pole of $S^2$}\label{sec:3}
We begin by considering the correlation function between a Wilson loop lying on a latitude of
$S^2$ and a chiral primary operator inserted at the north pole. 
In our coordinate system the north pole is $x_N=(0,0,1,0)$ and, as a 
consequence, the CPO operator assumes a very simple form because 
just two scalars ($\Phi^3,\Phi^4$) appear in its explicit expression. 
It is useful instead to write the Wilson loop through a generalized 
connection
\begin{equation}\label{wilsonpsi}
W[\mathcal{C}]=\frac{1}{N}\text{Tr}\mathcal{P}\exp{\oint d\tau 
\;\mathcal{A}(x(\tau))},
\end{equation}
where
\begin{equation}\label{connection1}
 \mathcal{A}(x(\tau))=(iA_\mu 
 \dot{x}^\mu+\sin^2{\theta}\,\Phi^3-\sin{\theta}\cos{\theta}(\sin{\tau}\,\Phi^2+\cos{\tau}\,\Phi^1)).
\end{equation}
Here $\theta$ is the latitude angle in standard polar coordinates and for symmetry reasons we will 
restrict its range to $[0,\pi/2]$: at $\theta=0$ 
the contour shrinks to the north pole while at $\theta=\pi/2$ we 
get the equator of $S^2$. The position on the latitude is parametrized 
by $\tau$, ranging from $0$ to $2\pi$, and we will denote, in the 
case of multiple integrations, $x(\tau_i)=x_i$, $\Phi^I(x_i)=\Phi_i^I$ 
and $\mathcal{A}(x_i)=\mathcal{A}_i$. The effective propagators entering the actual 
computation do not depend on the positions along the latitude and are the following
\begin{equation}
  \begin{split}\label{effprop1}
    \langle\mathcal{A}_i^{ab}\,\mathcal{A}_j^{cd}\rangle&=\frac{\lambda'}{16\pi^2}\frac{\delta^{ad}\delta^{bc}}{N},\\
    \langle\mathcal{A}_i^{ab}\,\Phi^{I\,cd}(x_N)\rangle&=\frac{\lambda'}{16\pi^2}\frac{\delta_{I3}}{1-\cos\theta}\frac{\delta^{ad}\delta^{bc}}{N},    
  \end{split}
\end{equation}
where $\lambda'=4\lambda A_1A_2/A^2=\lambda\sin^2\theta$ for
this loop.

In the following we will restrict our investigation to the case $J=2$ and at order $\lambda^2$. 
This choice will simplify our analysis and at two-loop level does 
not represent a real limitation: no new class of perturbative diagrams 
would enter the computation and the general case should be tamed 
by simple combinatorics.

\subsection{Ladder contribution I}
\begin{figure}[!h]
\centering
\includegraphics[width=4.6cm]{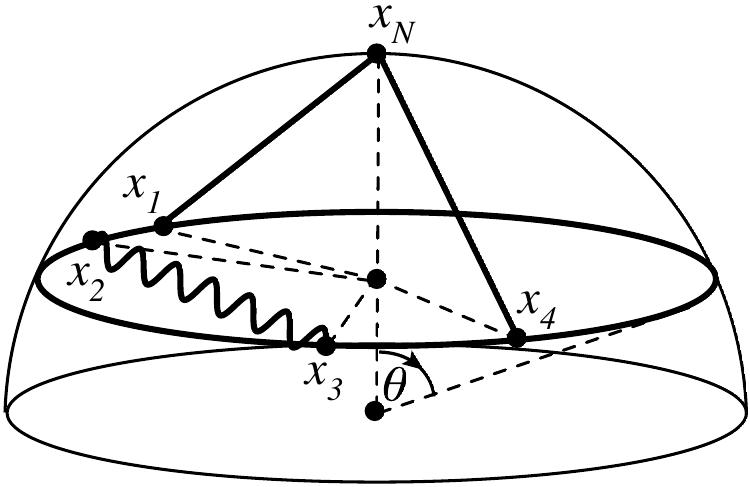}
\caption{Ladder diagram for latitude-north pole correlation function at order $\lambda^2$.}
\label{fig:latlad}
\end{figure}

Ladder diagrams are the easiest class of perturbative contributions to our correlation function: 
using the effective propagators \eqref{effprop1}, a straightforward calculation yelds
\begin{equation}\label{laddero2}
\langle W[\mathcal{C}]\mathcal{O}_2(x_N)\rangle_{\text{ladder}}=
\frac{1}{N}\frac{\lambda'^2}{192\sqrt{2}}\left(\frac{A_2}{A_1}\right),
\end{equation}
where $\frac{A_2}{A_1}=\cot^2\frac\theta 2$. 
Actually it is not difficult to derive the ladder contribution for  general $J$ and at 
any perturbative order (see appendix \ref{sec:appB}): in this case it corresponds to the matrix
model result \eqref{matrmodel}. Of course (\ref{laddero2}) particularizes this result.

\subsection{Interacting contributions I}

\begin{figure}[!h]
\centering
\subfigure[]{\includegraphics[width=4.6cm]{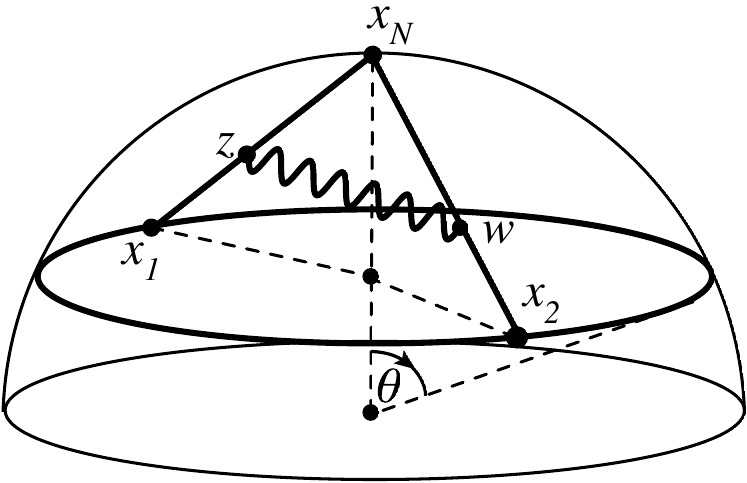}}\quad
\subfigure[]{\includegraphics[width=4.6cm]{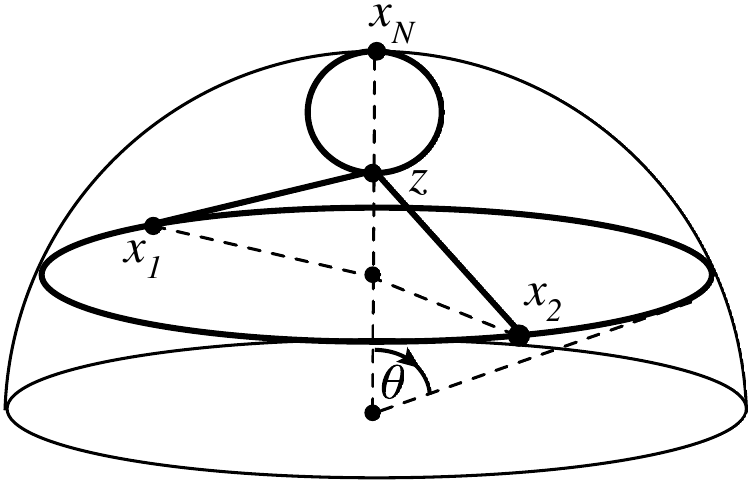}}\\
\subfigure[]{\includegraphics[width=4.6cm]{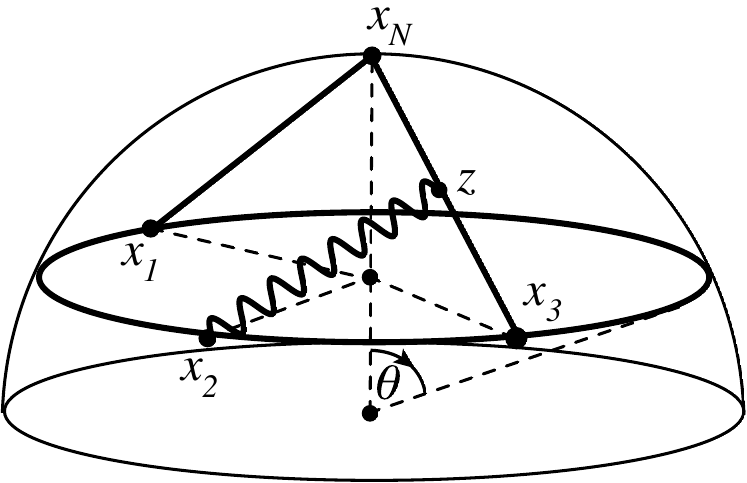}}\quad
\subfigure[]{\includegraphics[width=4.6cm]{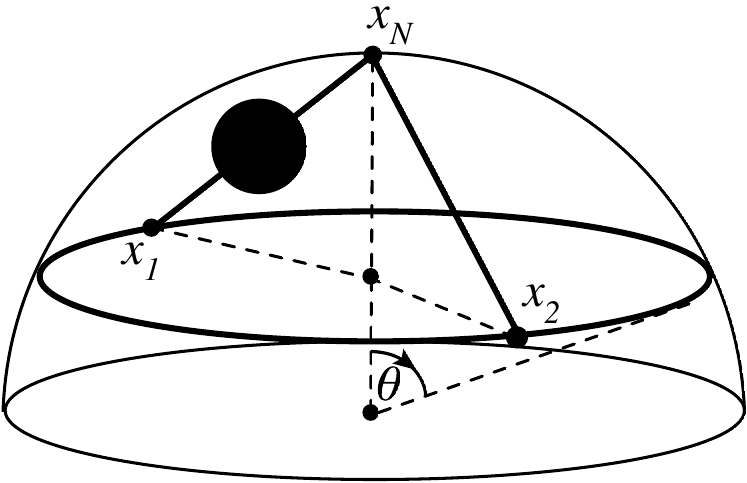}}
\caption{Diagrams with interaction vertices at order $\lambda^2$: (a) \textbf{H}-contribution, (b) \textbf{X}-contribution,
 (c) \textbf{IY}-contribution and (d) \textbf{O}-contribution; $z$ and $w$ denote the position of interaction vertices.}
\label{fig:latint}
\end{figure}

We discuss now the effect of interaction vertices to the correlation 
function at order $\lambda^2$: it is the crucial part of the computation. 
We expect indeed that their total contribution sums to zero since, for the particular configuration we are 
considering, ladder diagrams are enough to recover the matrix model 
expression, as shown by (\ref{laddero2}). The different interacting 
diagrams are grouped in four classes, denoted by \textbf{H}, \textbf{X}, \textbf{IY} and \textbf{O}, 
symbols that actually resemble their graphical form (see Figure (\ref{fig:latint})).

\paragraph{The H-contribution:}
We first consider the diagrams of type \textbf{H}: the interaction vertices 
are connected here by a 
gluon propagator and its form is
\small\begin{equation}
-\frac{2}{(2\pi)^2\lambda}\int_0^{2\pi} d\tau_1 \int_0^{\tau_1} d\tau_2\int d^4 z\;d^4 w\;D(z-w)\; 
\langle \text{Tr}{(\mathcal{A}_1\mathcal{A}_2)}\;\text{Tr}(\Phi_z^I\partial^\mu_z\Phi_z^I[\Phi_w^J,\partial^\mu_w\Phi_w^J])\;
\mathcal{O}_2(x_N)\rangle ,
\end{equation}\normalsize
where $D(x-y)=\frac{1}{(x-y)^2}$. Due to the explicit form of the 
CPO and of the Wilson loop we have non vanishing contributions 
from $I=J=3$. Defining the composite 
operator $\mathcal{O}_2(x_N)$ via point-splitting
\begin{equation}
(\Phi(x_N))^2=\lim_{\substack{y_1\rightarrow x_N \\ 
y_2\rightarrow x_N}}\Phi(y_1)\Phi(y_2),
\end{equation}
a straightforward manipulation leads to
\begin{equation}
  \begin{split}
\text{\textbf{H}}=&\frac{\lambda'^2}{2^2\sqrt{2}(2\pi)^8N}\int_0^{2\pi} d\tau_1 \int_0^{\tau_1} d\tau_2\int d^4 z\;d^4 w\,D(z-w)\\ 
&\quad\times\biggr[\biggr(\partial_{x_1}-\partial_{y_1}\biggr)\cdot\biggr(\partial_{x_2}-\partial_{y_2}\biggr)
D(z-y_1)D(w-y_2)D(z-x_1)D(w-x_2)\\
&\quad+\biggr(\partial_{x_2}-\partial_{y_1}\biggr)\cdot\biggr(\partial_{x_1}-\partial_{y_2}\biggr)
D(w-y_1)D(z-y_2)D(z-x_1)D(w-x_2)\biggr]\\
=&\frac{(2\pi)^2\lambda'^2}{2^2\sqrt{2}N}\int_0^{2\pi} d\tau_1 \int_0^{2\pi} d\tau_2
\;(\partial_{x_1}-\partial_{y_1})\cdot(\partial_{x_2}-\partial_{y_2})
\;\mathcal{H}(x_1,y_1;x_2,y_2).
  \end{split}
\end{equation}
Here we have used the symmetry $z\leftrightarrow w$, symmetrized 
the expression in the exchange $x_1\leftrightarrow x_2$ and defined
\begin{equation}
\mathcal{H}(x_1,x_2;x_3,x_4)=\int \frac{d^4 z\;d^4 
w}{(2\pi)^{10}}D(z-x_1)D(z-x_2)D(z-w)D(w-x_3)D(w-x_4).
\end{equation}
Taking advantage of the identity \cite{Beisert:2002bb}
\begin{equation}\begin{split}\label{qwe}
(\partial_{x_1}-&\partial_{y_1})\cdot(\partial_{x_2}-\partial_{y_2})
\;\mathcal{H}(x_1,y_1;x_2,y_2)\\
=&\frac{1}{(x_1-y_1)^2(x_2-y_2)^2}\biggr[
\mathcal{X}(x_1,y_1,x_2,y_2)\biggr((x_1-x_2)^2(y_1-y_2)^2-(x_1-y_2)^2(x_2-y_1)^2\biggr)\\
&+\frac{1}{(2\pi)^2}\biggr(\mathcal{G}(x_1;x_2,y_2)-\mathcal{G}(y_1;x_2,y_2)+
\mathcal{G}(x_2;x_1,y_1)-\mathcal{G}(y_2;x_1,y_1)\biggr)\biggr],
\end{split}\end{equation}
where
\begin{equation}\begin{split}\label{def}
\mathcal{G}(x_1;x_2,x_3)&=\mathcal{Y}(x_1,x_2,x_3)[(x_1-x_3)^2-(x_1-x_2)^2],\\
\mathcal{X}(x_1,x_2,x_3,x_4)&=\frac{1}{(2\pi)^8}\int\frac{d^4 
z}{(z-x_1)^2(z-x_2)^2(z-x_3)^2(z-x_4)^2},\\
\mathcal{Y}(x_1,x_2,x_3)&=\frac{1}{(2\pi)^6}\int\frac{d^4 
z}{(z-x_1)^2(z-x_2)^2(z-x_3)^2}\equiv \mathcal{I}_1(x_1-x_3,x_2-x_3),
\end{split}\end{equation}
and setting $y_1=y_2=x_N$ we arrive at
\begin{equation}\begin{split}
\text{\textbf{H}}=&-\frac{\lambda'^2}{2^2\sqrt{2}N}\oint d\tau_1 d\tau_2
\;\mathcal{X}(x_1,x_N,x_2,x_N)+\\
&+\frac{\lambda'^2}{2^2\sqrt{2}N}\frac{1}{1-\cos\theta}\oint d\tau_1 d\tau_2
\;\biggr[\mathcal{I}_1(x_1-x_N,x_2-x_N)+\mathcal{I}_1(0,x_2-x_N)\biggr]\\
&-\frac{\lambda'^2}{2^3\sqrt{2}N}\frac{1}{(1-\cos\theta)^2}\oint d\tau_1 d\tau_2
\;\;\mathcal{I}_1(x_1-x_2,x_N-x_2)\;(x_1-x_2)^2.
\end{split}\end{equation}

\paragraph{The X-contribution:}
The \textbf{X} diagram comes entirely from the four-point scalar vertex
\begin{equation}\label{prob}\small
\frac{1}{2\lambda}\int_0^{2\pi} d\tau_1 \int_0^{\tau_1} d\tau_2\int d^4 z\; 
\langle 
\text{Tr}{(\mathcal{A}_1\mathcal{A}_2)}\;\text{Tr}([\Phi^I_z,\Phi^J_z]^2)\;\mathcal{O}_2(x_N)\rangle.
\end{equation}
The only non vanishing terms arise from $\mathcal{A}_i=\Phi_i^3$ and $I,J=3,4$, giving directly
\begin{equation}
\text{\textbf{X}}=\frac{(2\pi)^2\lambda'^2}{2^2\sqrt{2}N}\int_0^{2\pi} d\tau_1 \int_0^{2\pi} d\tau_2
\;\mathcal{X}(x_N,x_N,x_1,x_2).
\end{equation}

\paragraph{The IY-contribution:}
We examine the most elaborate part of the two-loop computation, 
involving the presence of three distinct contour integrations
\begin{equation}\label{prob}\small
\frac{i2}{3\lambda}\oint d\tau_1 d\tau_2 d\tau_3 \;\eta(\tau_1,\tau_2,\tau_3)\int d^4 z\; \langle 
\text{Tr}{(\mathcal{A}_1\mathcal{A}_2\mathcal{A}_3)}\;\text{Tr}(\partial^\nu_z\Phi_z[A^\nu_z,\Phi_z])\;\mathcal{O}_2(x_N)\rangle,
\end{equation}
where to take into account the appropriate ordering we have defined
\begin{equation}\label{eta}
\eta(\tau_1,\tau_2,\tau_3)=\theta(\tau_1-\tau_2)\theta(\tau_2-\tau_3)+\text{cyclic permutations}.
\end{equation}
Computing the contractions we get
\begin{equation}\begin{split}
\text{\textbf{IY}}=&-\frac{\lambda'^2}{2^2\sqrt{2}N}\oint d\tau_1 d\tau_2 d
\tau_3 \,\epsilon(\tau_1,\tau_2,\tau_3)\;\dot{x}_2^\mu\, D(x_1-x_N)\\
&\times\int \frac{d^4 z}{(2\pi)^6}
\,D(x_2-z)\biggr[D(x_N-z)\partial_z D(x_3-z)-\partial_z 
D(x_N-z)D(x_3-z)\biggr],
\end{split}\end{equation}
with
\begin{equation}\label{epsilon}
\epsilon(\tau_1,\tau_2,\tau_3)=\eta(\tau_1,\tau_2,\tau_3)-\eta(\tau_2,\tau_1,\tau_3).
\end{equation}
Performing an integration by parts, we can rewrite \textbf{IY} as
\begin{equation}\begin{split}\label{formulone}
\text{\textbf{IY}}=&\frac{\lambda'^2}{2^2\sqrt{2}N}\oint d\tau_1 d\tau_2 d
\tau_3 \,\epsilon(\tau_1,\tau_2,\tau_3)\;\dot{x}_2^\mu\, 
D(x_1-x_N)(2\partial_3+\partial_2)\mathcal{Y}(x_2,x_3,x_N)\\
=&\frac{\lambda'^2}{2^2\sqrt{2}N}\frac{1}{1-\cos\theta}\biggr\{\oint d\tau_1d
\tau_3\,\biggr[\mathcal{I}_1(x_1-x_N,x_3-x_N)-\mathcal{I}_1(0,x_3-x_N)\biggr]\\
&\qquad\qquad\qquad\qquad\qquad+\oint d\tau_1 d\tau_2 d
\tau_3 \,\epsilon(\tau_1,\tau_2,\tau_3) 
\;\dot{x}_2^\mu\,\partial_3\,\mathcal{I}_1(x_3-x_N,x_2-x_N)\biggr\},
\end{split}\end{equation}
where we have used
\begin{equation}\label{depsilon}
\frac{\partial}{\partial 
\tau_2}\epsilon(\tau_1,\tau_2,\tau_3)=2(\delta(\tau_2-\tau_3)-\delta(\tau_1-\tau_2)).
\end{equation}
The triple integral can be massaged exploting the trivial identity
\begin{equation}\label{pucciid}
\frac{\lambda'^2}{2^2\sqrt{2}N}\frac{1}{1-\cos\theta}\oint d\tau_1 d\tau_2 d
\tau_3 \,\frac{d}{d\tau_2}\biggr[\epsilon(\tau_1,\tau_2,\tau_3) 
\;\mathcal{I}_2(x_3-x_N,x_2-x_N)\biggr]=0,
\end{equation}
where the function $\mathcal{I}_2$ is defined in the appendix \ref{sec:appA}. 
Upon subtracting \eqref{pucciid} to \eqref{formulone} we obtain
\begin{equation}
  \begin{split}\label{iyvmu}
\text{\textbf{IY}}=\frac{\lambda'^2}{2^2\sqrt{2}N}\frac{1}{1-\cos\theta}\biggr\{&\oint d\tau_1d
\tau_3\,\biggr[\mathcal{I}_1(x_1-x_N,x_3-x_N)-\mathcal{I}_1(0,x_3-x_N)\biggr]\\
-&2\oint d\tau_1d
\tau_3\,\biggr[\mathcal{I}_2(x_3-x_N,x_3-x_N)-\mathcal{I}_2(x_3-x_N,x_1-x_N)\biggr]\\
+&\oint d\tau_1 d\tau_2 d
\tau_3 \,\epsilon(\tau_1,\tau_2,\tau_3) 
\;\dot{x}_2^\mu\,V_\mu(x_3-x_N,x_2-x_N)\biggr\},
\end{split}\end{equation}
with
\begin{equation}\label{Vumu}
V^\mu(x,y)\equiv 
\partial_x^\mu\,\mathcal{I}_1(x,y)-\partial_y^\mu\,\mathcal{I}_2(x,y).
\end{equation}
With the help of equation (\ref{vuide}) we find
\small\begin{equation}\begin{split}
\dot{x}_2^\mu\,V_\mu(x_3-x_N,x_2-x_N)=
-&\frac{1}{32\pi^4(x_3-x_N)^2}\frac{d}{d t_2}\biggr[
\mathrm{Li}_2\left(1-\frac{(x_3-x_2)^2}{(x_2-x_N)^2}\right)\\
&\qquad\qquad\qquad+\frac12\log^2\left[\frac{(x_3-x_2)^2}{(x_2-x_N)^2}\right]
-\frac{1}{2} \log^2\left[\frac{(x_3-x_2)^2}{(x_3-x_N)^2}\right]\biggr].
\end{split}\end{equation}
\normalsize
Inserting this result into (\ref{iyvmu}) and integrating by parts we arrive at 
the final expression
\begin{equation}
  \begin{split}
\text{\textbf{IY}}=\frac{\lambda'^2}{2^2\sqrt{2}N}&\frac{1}{1-\cos\theta}\biggr\{\oint d\tau_1d
\tau_3\,\biggr[\mathcal{I}_1(x_1-x_N,x_3-x_N)-\mathcal{I}_1(0,x_3-x_N)\biggr]\\
-&2\oint d\tau_1d
\tau_3\,\biggr[\mathcal{I}_2(x_3-x_N,x_3-x_N)-\mathcal{I}_2(x_3-x_N,x_1-x_N)\biggr]\\
-&\frac{1}{2^6\pi^4}\frac{1}{1-\cos\theta}\oint d\tau_1 d\tau_3 \,
\Biggr[\mathrm{Li}_2\left(1-\frac{\sin^2\theta}{1-\cos\theta}(1-\cos\tau_{31})\right)-\frac{\pi^2}{6}\Biggr]
\biggr\},
\end{split}\end{equation}
where $\tau_{ij}=\tau_i-\tau_j$.

\paragraph{The O-contribution:}
The last interacting diagram comes from the self-energy of the scalar 
propagator: borrowing directly the result from \cite{Erickson:2000af}, we write down
\begin{equation}
\text{\textbf{O}}=-\frac{\lambda'^2}{2\sqrt{2}N}\frac{1}{1-\cos\theta}\oint d\tau_1 
d\tau_2\;\,\mathcal{I}_1(0,x_2-x_N).
\end{equation}

\subsection{Summing up interactions I}
Adding up the contributions of all the interacting diagrams 
we obtain:
\begin{equation}
  \begin{split}\label{sommaint1}
\langle W[\mathcal{C}]&\mathcal{O}_2(x_N)\rangle_{int}=\text{\textbf{H}}+\text{\textbf{O}}+\text{\textbf{X}}+\text{\textbf{IY}}=\\
&-\frac{\lambda'^2}{2^3\sqrt{2}N}\frac{1}{(1-\cos\theta)^2}\oint d\tau_1\,d\tau_2
\;\;\mathcal{I}_1(x_1-x_2,x_N-x_2)\;(x_1-x_2)^2\\
&+\frac{\lambda'^2}{2\sqrt{2}N}\frac{1}{1-\cos\theta}\oint d\tau_1\,d\tau_2
\;\biggr[\mathcal{I}_1(x_1-x_N,x_2-x_N)+\mathcal{I}_2(x_2-x_N,x_1-x_N)\biggr]\\
&-\frac{\lambda'^2}{2\sqrt{2}N}\frac{1}{1-\cos\theta}\oint d\tau_1\,d\tau_2
\;\biggr[\mathcal{I}_2(x_2-x_N,x_2-x_N)+\mathcal{I}_1(0,x_2-x_N)\biggr]\\
&-\frac{\lambda'^2}{2^8\pi^4\sqrt{2}N}\frac{1}{(1-\cos\theta)^2}\oint d\tau_1 d
\tau_2\Biggr[\mathrm{Li}_2\left(1-\frac{\sin^2\theta}{1-\cos\theta}(1-\cos\tau_{21})\right)-\frac{\pi^2}{6}\Biggr].
\end{split}
\end{equation}
Remarkably, no triple contour integration is present in this final 
expression. The integrals in \eqref{sommaint1}, denoted by $P_{1,2,3,4}$, are evaluated in 
appendix \ref{sec:appC}. Using these results we find
\begin{equation}
  \begin{split}
\langle W[\mathcal{C}]\mathcal{O}_2(x_N)\rangle_{int}=
-\frac{\lambda'^2}{2^3\sqrt{2}N}\frac{1}{(1-\cos\theta)^2}\left[P_1-
\frac{1-\cos\theta}{4}(P_2-P_3)+\frac{P_4}{2^5\pi^4}-\frac{1}{3\cdot 
2^4}\right]=0
\end{split}
\end{equation}
for any $\theta$, as expected.
We confirm therefore that at order $\lambda^2$, the correlator of the latitude Wilson loop with $\mathcal{O}_2$
at the north-pole is
\begin{equation}
\langle W[\mathcal{C}]\mathcal{O}_2(x_N)\rangle=
\langle W[\mathcal{C}]\mathcal{O}_2(x_N)\rangle_{\text{ladder}}=
\frac{1}{N}\frac{\lambda'^2}{192\sqrt{2}}\left(\frac{A_2}{A_1}\right).
\end{equation}

\section{Perturbative computations II: equator Wilson loop with an operator
insertion at an arbitrary point of $S^2$}\label{sec:4}

In this section we consider the correlation function of a Wilson loop
shaped on the equator of $S^2$
and the CPO operator (\ref{oj}) inserted on the sphere at the point $x_{\mathcal{O}}=(\sin\phi,0,\cos\phi)$ 
(one of the coordinates can be taken to zero by symmetry reasons). Without loss of
generality, we also assume that the operator is located in the north
hemisphere, and we consider $0\leq \phi\leq \pi/2$. Thus the CPO depends on the three scalars $\Phi^I$, with
$I=1,3,4$ and the Wilson loop is written as an integral of the generalized connection as in (\ref{wilsonpsi}) 
with
\begin{equation}
 \mathcal{A}(x(\tau))=(iA_\mu \dot{x}^\mu+\Phi^3).
\end{equation}
The effective propagators are now the following:
\begin{equation}
  \begin{split}\label{effprop2}
    \langle\mathcal{A}_i^{ab}\;\mathcal{A}_j^{cd}\rangle&=\frac{\lambda}{16\pi^2}\frac{\delta^{ad}\delta^{bc}}{N},\\
    \langle\mathcal{A}_i^{ab}\;\Phi^{I\,cd}(x_\mathcal{O})\rangle&=\frac{\lambda}{16\pi^2}f(\tau_i)\cos\phi\frac{\delta^{ad}\delta^{bc}}{N}\, \delta_{I 3},    
  \end{split}
\end{equation}
where
\begin{equation}\label{ftau}
f(\tau_i)=\frac{1}{1-\sin\phi\cos \tau_i}.
\end{equation}
Notice that a new and relevant feature appears in this case: effective
propagators connecting the CPO and the Wilson loop depend explicitly on the integration parameters $\tau_i$. 
As we will see soon, this aspect complicates considerably the computations and, crucially,
destroys the naive matrix model picture based on summing up ladder diagrams
and neglecting interaction vertices. Unfortunately we will not able to
perform all the calculations analytically and we will resort to numerical
integration for one particular contribution. We again limit ourselves to
CPO with $J=2$.

\subsection{Ladder contribution II}
\begin{figure}[!h]
\centering
\includegraphics[width=4.6cm]{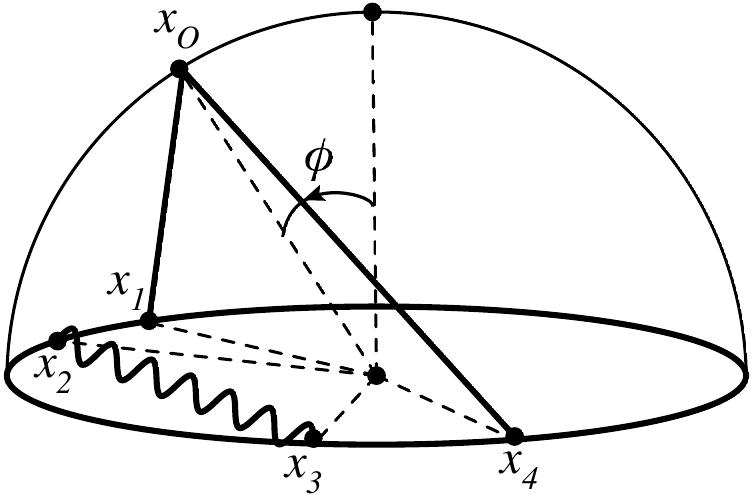}
\caption{Ladder diagram for equator-arbitrary point correlation function at order $\lambda^2$.}
\label{fig:eqlad}
\end{figure}

At the order $\lambda^2$, the ladder contribution arises from
\begin{equation}
\langle W[\mathcal{C}]\mathcal{O}_2(x_{\mathcal{O}})\rangle_{\text{ladder}} 
=\frac{1}{N}\int_0^{2\pi}d\tau_1\int_0^{\tau_{1}}d\tau_{2}\int_0^{\tau_{2}}d\tau_{3}\int_0^{\tau_{3}}d\tau_{4}
\,\langle \text{Tr}(\mathcal{A}_1\mathcal{A}_2\mathcal{A}_3\mathcal{A}_{4})\,\mathcal{O}_2(x_{\mathcal{O}})\rangle.
\end{equation} 
By performing the contractions and using \eqref{effprop2}, we find
\small\begin{equation}\label{o2xop2}\begin{split}
\frac{\lambda^2\cos^2\phi}{2^9\sqrt{2}\pi^4 N}\int_0^{2\pi}d\tau_1...\int_0^{\tau_{3}}d\tau_{4}
\,\biggr[f(\tau_1)f(\tau_4)+f(\tau_1)f(\tau_2)+f(\tau_2)f(\tau_3)+f(\tau_3)f(\tau_4)\biggr].
\end{split}
\end{equation}\normalsize 
By simply changing the integration order, we can evaluate two integrals,
ending up with
\small\begin{equation}\label{o2xop3}\begin{split}
&\frac{\lambda^2\cos^2\phi}{2^9\sqrt{2}\pi^4 N}
\int_0^{2\pi}d\tau_1\,f(\tau_1)\int_0^{\tau_{1}}d\tau_{2}\,f(\tau_2)
\biggr[(\tau_1-\tau_2)^2+2\pi^2-2\pi(\tau_1-\tau_2)\biggr]\\
=&\frac{\lambda^2\cos^2\phi}{2^9\sqrt{2}\pi^4 N}
\biggr[\mathcal{J}_2\mathcal{J}_0-\mathcal{J}_1^2+\pi^2\mathcal{J}_0^2-2\pi\mathcal{J}_1\mathcal{J}_0+4\pi\tilde{\mathcal{J}}\biggr],
\end{split}
\end{equation}\normalsize
where $\mathcal{J}_n$ e $\tilde{\mathcal{J}}$ are defined and computed in
appendix \ref{sec:appD}. Using these results we get
\small\begin{equation}\begin{split}\label{ladder2}
\langle W[\mathcal{C}]&\mathcal{O}_2(x_{\mathcal{O}})\rangle_{\text{ladder}}=\\
=&\frac{\lambda^2}{192\sqrt{2}N}
-\frac{\lambda^2}{2^6\sqrt{2}\pi^2}\biggr[\log\left(\frac{2\sigma}{1+\sigma}\right)^2+
\log\left(\frac{1+\sigma}{2}\right)^2+2\text{Li}_2\left(\frac{1-\sigma}{2}\right)+2\text{Li}_2\left(\frac{\sigma-1}{2\sigma}\right)\biggr],
\end{split}\end{equation}\normalsize
where $\sigma=\sqrt{\frac{1+\sin\phi}{1-\sin\phi}}$.

The first term in the above expression already gives the matrix model result, 
i.e. the second order term in the expansion of Bessel $I_2(\sqrt\lambda)$. Therefore
the remaining term
\begin{equation}\label{L}
\text{\textbf{L}}\equiv -\frac{\lambda^2}{2^6\sqrt{2}\pi^2}\biggr[\log\left(\frac{2\sigma}{1+\sigma}\right)^2+
\log\left(\frac{1+\sigma}{2}\right)^2+2\text{Li}_2\left(\frac{1-\sigma}{2}\right)+2\text{Li}_2\left(\frac{\sigma-1}{2\sigma}\right)\biggr]
\end{equation}
should cancel the interacting contributions.

\subsection{Interacting contributions II}

We attempt here the computation of the diagrams containing interaction
vertices: due to the asymmetry of our configuration we will not be able to
obtain an expression only in terms of double-integrals, as in the previous
case. We have truly to face triple contour integrations, and moreover part of the job
must be done numerically.
\begin{figure}[!h]
\centering
\subfigure[]{\includegraphics[width=4.6cm]{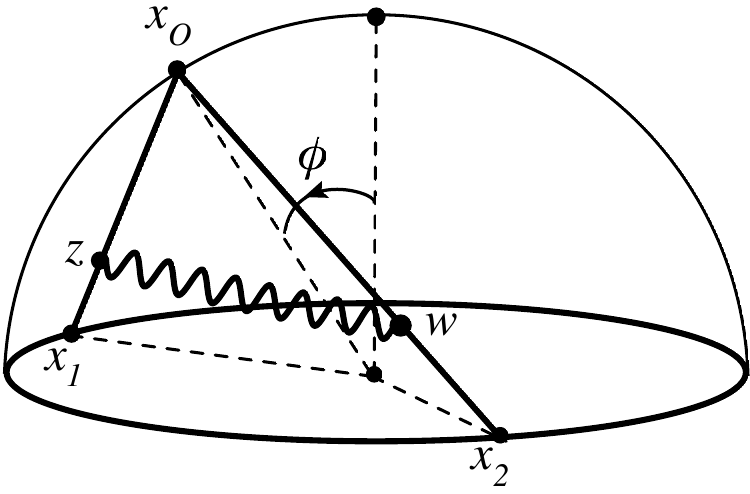}}\quad
\subfigure[]{\includegraphics[width=4.6cm]{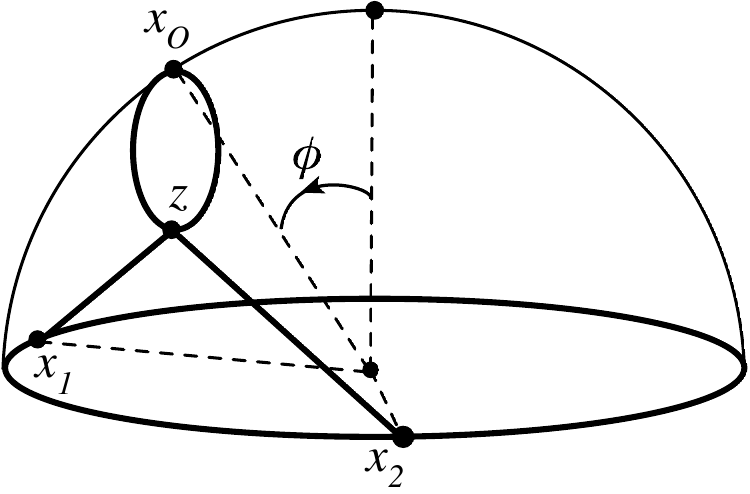}}\\
\subfigure[]{\includegraphics[width=4.6cm]{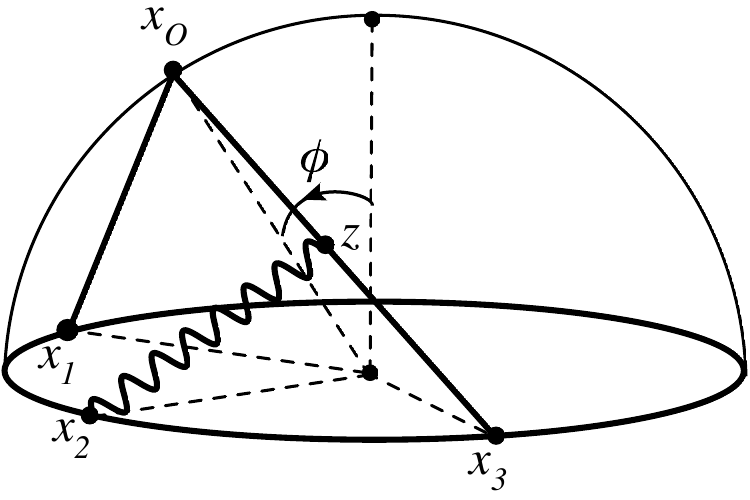}}\quad
\subfigure[]{\includegraphics[width=4.6cm]{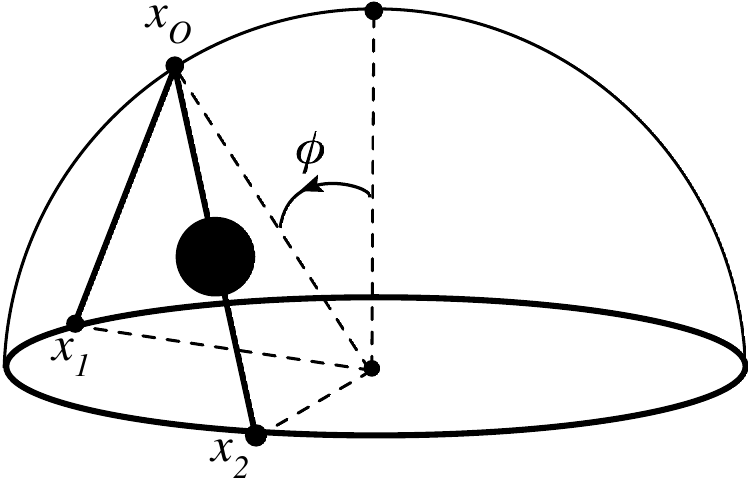}}
\caption{Diagrams with interaction vertices at order $\lambda^2$: (a) \textbf{H}-contribution, (b) \textbf{X}-contribution,
 (c) \textbf{IY}-contribution and (d) \textbf{O}-contribution; $z$ and $w$ denote the position of interaction vertices.}
\label{fig:eqint}
\end{figure}

\paragraph{The H, X and O contributions:}
The procedure to evaluate the diagrams \textbf{H}, \textbf{O} e \textbf{X}
is very similar to the previous case. Their structure remains basically
unchanged, the only relevant difference being the appearance of contour dependent 
propagators. We get
\begin{align}\begin{split}
\text{\textbf{H}}=&-\frac{(2\pi)^2\lambda^2\cos^2\phi}{2^2\sqrt{2}N}\oint d\tau_1 d\tau_2
\;\mathcal{X}(x_1,x_{\mathcal{O}},x_2,x_{\mathcal{O}})\\
&+\frac{\lambda^2\cos^2\phi}{2^2\sqrt{2}N}\int_0^{2\pi} d\tau_1\,f(\tau_1)\int_0^{2\pi} d\tau_2
\;\biggr[\mathcal{I}_1(x_1-x_{\mathcal{O}},x_2-x_{\mathcal{O}})+\mathcal{I}_1(0,x_2-x_{\mathcal{O}})\biggr]\\
&-\frac{\lambda^2\cos^2\phi}{2^3\sqrt{2}N}\int_0^{2\pi} 
d\tau_1\,f(\tau_1)\int_0^{2\pi}d\tau_2\,f(\tau_2)
\;\;\mathcal{I}_1(x_1-x_{2},x_{\mathcal{O}}-x_2)\;(x_1-x_2)^2,
\end{split}\\[8pt]
\text{\textbf{X}}=&\frac{(2\pi)^2\lambda^2\cos^2\phi}{2^2\sqrt{2}N}\oint d\tau_1 d\tau_2
\;\mathcal{X}(x_{\mathcal{O}},x_{\mathcal{O}},x_1,x_2),\\
\text{\textbf{O}}=&-\frac{\lambda^2\cos^2\phi}{2\sqrt{2}N}\int_0^{2\pi} d\tau_1\,f(\tau_1)\int_0^{2\pi} d\tau_2
\;\,\mathcal{I}_1(0,x_2-x_{\mathcal{O}}),
\end{align}
with the functions $\mathcal{X}$ e $\mathcal{I}_1$ defined in (\ref{def}).

\paragraph{The IY-contribution:}
We have seen in the previous section, that the \textbf{IY}  diagram contains 
triple integrations along the circuit. In that case we have been able,
through some judicious manipulation, to reduce the problem to double
integrals. Now, with the CPO operator in an arbitrary position on the sphere, this technique works only partially.

Repeating the same steps as in section \ref{sec:3} we arrive at the expression
\begin{equation}\begin{split}
\text{\textbf{IY}}=\frac{\lambda^2\cos^2\phi}{2^2\sqrt{2}N}\oint d\tau_1\,f(\tau_1)\biggr\{&\oint d
\tau_3\,\biggr[\mathcal{I}_1(x_1-x_{\mathcal{O}},x_3-x_{\mathcal{O}})-\mathcal{I}_1(0,x_3-x_{\mathcal{O}})\biggr]\\
&+\oint d\tau_2 d
\tau_3 \,\epsilon(\tau_1,\tau_2,\tau_3) 
\;\dot{x}_2^\mu\,\partial_3\,\mathcal{I}_1(x_3-x_{\mathcal{O}},x_2-x_{\mathcal{O}})\biggr\}.
\end{split}\end{equation}
We can still massage the triple integral using an identity similar to
(\ref{pucciid}), obtaining
\begin{equation}\begin{split}
\text{\textbf{IY}}=\frac{\lambda^2\cos^2\phi}{2^2\sqrt{2}N}\oint d\tau_1\,&f(\tau_1)\biggr\{\oint d
\tau_3\,\biggr[\mathcal{I}_1(x_1-x_{\mathcal{O}},x_3-x_{\mathcal{O}})-\mathcal{I}_1(0,x_3-x_{\mathcal{O}})\biggr]\\
&-2\oint d\tau_3 
\biggr[\mathcal{I}_2(x_3-x_{\mathcal{O}},x_3-x_{\mathcal{O}})-\mathcal{I}_2(x_3-x_{\mathcal{O}},x_1-x_{\mathcal{O}})\biggr]\\
&+\oint d\tau_2 d
\tau_3 \,\epsilon(\tau_1,\tau_2,\tau_3) 
\;\dot{x}_2^\mu\,V_\mu(x_3-x_{\mathcal{O}},x_2-x_{\mathcal{O}})\biggr\}.
\end{split}\end{equation}
Using again \eqref{vuide}, we end up with
\begin{equation}\begin{split}
\text{\textbf{IY}}=\frac{\lambda^2\cos^2\phi}{2^2\sqrt{2}N}\oint &d\tau_1\,f(\tau_1)\biggr\{\oint d
\tau_3\,\biggr[\mathcal{I}_1(x_1-x_{\mathcal{O}},x_3-x_{\mathcal{O}})-\mathcal{I}_1(0,x_3-x_{\mathcal{O}})\biggr]\\
&-2\oint d\tau_3 
\biggr[\mathcal{I}_2(x_3-x_{\mathcal{O}},x_3-x_{\mathcal{O}})-\mathcal{I}_2(x_3-x_{\mathcal{O}},x_1-x_{\mathcal{O}})\biggr]\\
&-\frac{1}{2^6\pi^4}\oint d\tau_3\,f(\tau_3)
\Biggr[\mathrm{Li}_2\biggr(1-(1-\cos\tau_{31})f(\tau_1)\biggr)-\frac{\pi^2}{6}\Biggr]\\
&+\frac{1}{2^7\pi^4}\oint d\tau_2 d
\tau_3\,\epsilon(\tau_1,\tau_2,\tau_3)\,f(\tau_3) \,\cot{\left(\frac{\tau_{32}}{2}\right)}\log\left(\frac{f(\tau_3)}{f(\tau_2)}\right)
\biggr\}.
\end{split}\end{equation} 
Unfortunately, in this case the awkward triple integral cannot be avoided.

\subsection{Summing up interactions II}

\begin{figure}[!ht]
    \centering
    \includegraphics[width=11cm]{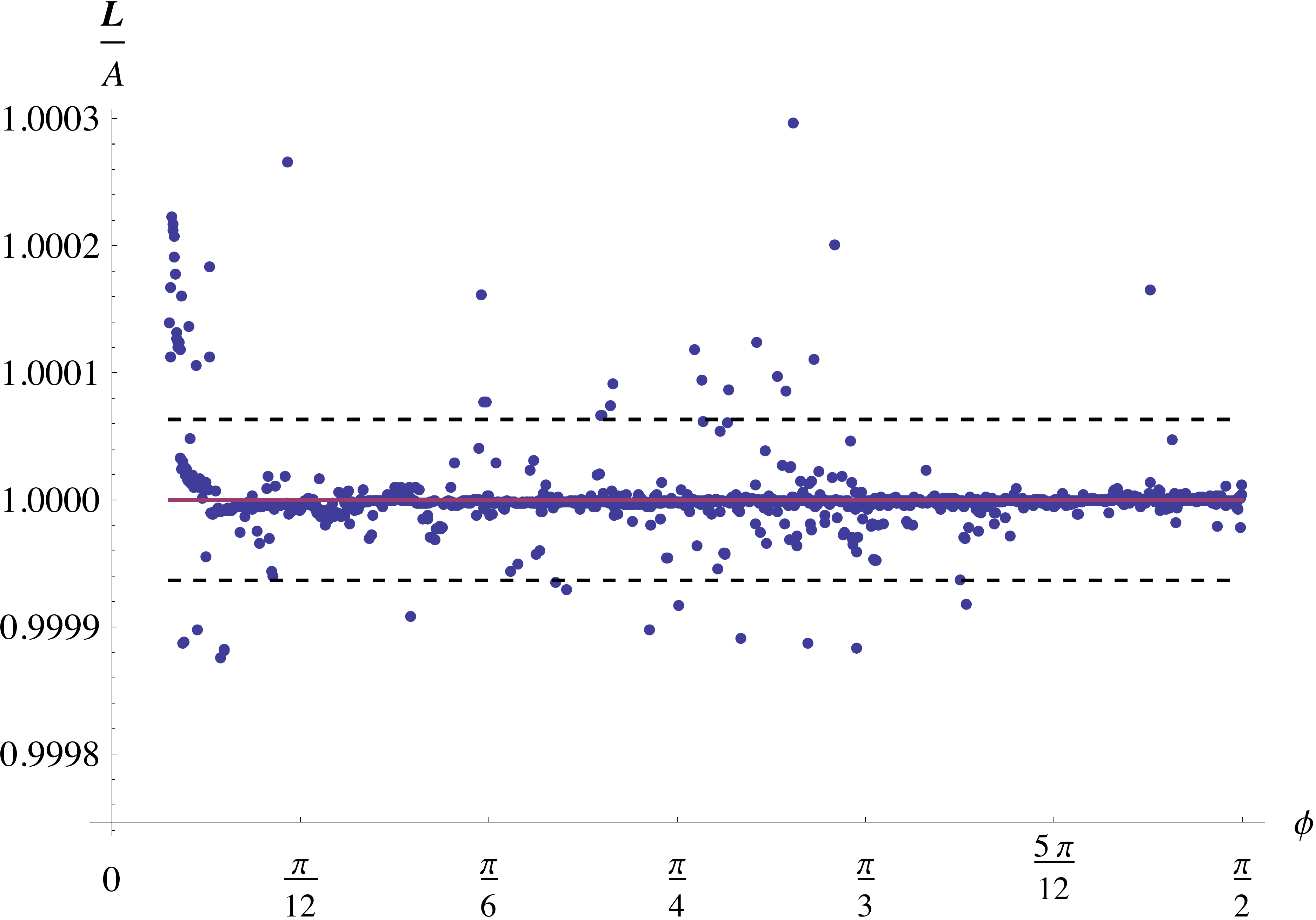}
    \caption{Numerical evaluation of the ratio $\frac{\text{\textbf{L}}}{\text{A}}$ as a function
    of $\phi$ obtained with Wolfram Mathematica routine NIntegrate. The red line is the linear fit of the data
    and the dashed lines are the upper and lower limit of the Confidence Interval.}\label{fig:rapp}
\end{figure}

The evaluation of the whole interacting contributions requires some care:
first of all let us collect the different diagrams
\begin{equation}
  \begin{split}\label{tutto2}
\langle\,W[\mathcal{C}]&\mathcal{O}_2(x_{\mathcal{O}})\,\rangle_{\text{int}}=\text{\textbf{H}}+\text{\textbf{O}}+\text{\textbf{X}}+\text{\textbf{IY}}=\\
&-\frac{\lambda^2\cos^2\phi}{2^3\sqrt{2}N}\oint
d\tau_1 d\tau_2\,f(\tau_1)\,f(\tau_2)
\;\;\mathcal{I}_1(x_1-x_{2},x_{\mathcal{O}}-x_2)\;(x_1-x_2)^2\\
&+\frac{\lambda^2\cos^2\phi}{2\sqrt{2}N}\oint d\tau_1 d\tau_2\,f(\tau_1)
\;\biggr[\mathcal{I}_1(x_1-x_{\mathcal{O}},x_2-x_{\mathcal{O}})+\mathcal{I}_2(x_2-x_{\mathcal{O}},x_1-x_{\mathcal{O}})\biggr]\\
&-\frac{\lambda^2\cos^2\phi}{2^8\pi^4\sqrt{2}N}\oint d\tau_1 
d\tau_2\,f(\tau_1)\,f(\tau_2)
\Biggr[\mathrm{Li}_2\biggr(1-(1-\cos\tau_{21})f(\tau_1)\biggr)-\frac{\pi^2}{6}\Biggr]\\
&-\frac{\lambda^2\cos^2\phi}{2\sqrt{2}N}\oint d\tau_1 d\tau_2\,f(\tau_1)
\;\biggr[\mathcal{I}_2(x_2-x_{\mathcal{O}},x_2-x_{\mathcal{O}})+\mathcal{I}_1(0,x_2-x_{\mathcal{O}})\biggr]\\
&+\frac{\lambda^2\cos^2\phi}{2^9\pi^4\sqrt{2}N}\oint d\tau_1 d\tau_2 d
\tau_3\,\epsilon(\tau_1,\tau_2,\tau_3)\,f(\tau_1)\,f(\tau_3) \,\cot{\left(\frac{\tau_{32}}{2}\right)}\log\left(\frac{f(\tau_3)}{f(\tau_2)}\right).    
  \end{split}
\end{equation}
Using the definitions in appendix \ref{sec:appA} we can simplify this expression noticing that
\begin{equation}\begin{split}
\biggr[\mathcal{I}_2(x_2-x_{\mathcal{O}},x_2-x_{\mathcal{O}})+\mathcal{I}_1(0,x_2-x_{\mathcal{O}})\biggr]=&-\lim_{\epsilon\rightarrow 0}
\frac{\csc (\pi  \epsilon ) (\Gamma (\epsilon )-2 \Gamma (1-\epsilon ) \Gamma (2 \epsilon ))}{128\pi^{3+2\epsilon}[(x_2-x_{\mathcal{O}})^2]^{1+2\epsilon} \Gamma (1-\epsilon )}\\
=&\frac{1}{2^7\pi^4}f(\tau_2)\frac{\pi^2}{6}.
\end{split}\end{equation}
Then we rewrite \eqref{tutto2} as a sum of two terms
\begin{equation}
  \begin{split}\label{somma2}
\langle\,W[\mathcal{C}]&\mathcal{O}_2(x_{\mathcal{O}})\,\rangle_{\text{int}}=\\
&\begin{rcases}
&-\frac{\lambda^2\cos^2\phi}{2^3\sqrt{2}N}\oint
d\tau_1 d\tau_2\,f(\tau_1)\,f(\tau_2)
\;\;\mathcal{I}_1(x_1-x_{2},x_{\mathcal{O}}-x_2)\;(x_1-x_2)^2\\
&+\frac{\lambda^2\cos^2\phi}{2\sqrt{2}N}\oint d\tau_1 d\tau_2\,f(\tau_1)
\;\biggr[\mathcal{I}_1(x_1-x_{\mathcal{O}},x_2-x_{\mathcal{O}})+\mathcal{I}_2(x_2-x_{\mathcal{O}},x_1-x_{\mathcal{O}})\biggr]\\
&-\frac{\lambda^2\cos^2\phi}{2^8\pi^4\sqrt{2}N}\oint d\tau_1 
d\tau_2\,f(\tau_1)\,f(\tau_2)
\Biggr[\mathrm{Li}_2\biggr(1-(1-\cos\tau_{21})f(\tau_1)\biggr)\Biggr]
\end{rcases}\text{A}\\
&\;\,+\frac{\lambda^2\cos^2\phi}{2^9\pi^4\sqrt{2}N}\oint d\tau_1 d\tau_2 d
\tau_3\,\epsilon(\tau_1,\tau_2,\tau_3)\,f(\tau_1)\,f(\tau_3) \,\cot{\left(\frac{\tau_{32}}{2}\right)}\log\left(\frac{f(\tau_3)}{f(\tau_2)}\right).\qquad\quad\text{B}    
  \end{split}
\end{equation}
The integrals A and B are computed in appendix \ref{sec:appE}.
In particular B has been evaluated analytically obtaining 
$B=-2\text{\textbf{L}}$, where $\text{\textbf{L}}$ is given in \eqref{L}. The term A has
been calculated numerically for different values of the angle $\phi$.
In Figure \ref{fig:rapp} we plot the ratio $\frac{\text{\textbf{L}}}{\text{A}}$ 
and from this analysis we conclude that $\text{A}=\text{\textbf{L}}$
\footnote{For $\phi < \frac{\pi}{32}$ the value of A is much less than its error, while 
\text{\textbf{L}} is an analytic quantity. Thus in Figure~\ref{fig:rapp} we drop the points 
in that range.}. 

The contribution of the interacting diagrams is therefore
\begin{equation}
\langle\,W[\mathcal{C}]\mathcal{O}_2(x_{\mathcal{O}})\,\rangle_{\text{int}}=-\text{\textbf{L}}.
\end{equation}
Finally, summing up the interacting and the ladder contributions, we obtain 
\begin{equation}
\langle\,W[\mathcal{C}]\mathcal{O}_2(x_{\mathcal{O}})\,\rangle=
\langle\,W[\mathcal{C}]\mathcal{O}_2(x_{\mathcal{O}})\,\rangle_{\text{int}}+
\langle\,W[\mathcal{C}]\mathcal{O}_2(x_{\mathcal{O}})\,\rangle_{\text{ladder}}=
\frac{\lambda^2}{192\sqrt{2}N}
\end{equation}
that perfectly fits into the localization result.

\newpage\clearpage
\appendix

\section{The integrals $\mathbf{\mathcal{I}_1(x,y)}$ and $\mathbf{\mathcal{I}_2(x,y)}$}\label{sec:appA}
The integral  $\mathcal{I}_1(x,y)$, defined in \eqref{def}, has been
evaluated \cite{Bassetto:2008yf} in momentum space representation 
and using dimensional regularization ($\omega=2+\epsilon$)
\begin{equation}
\label{inizio}
\begin{split}
\mathcal{I}_1(x,y)&\equiv
\int \frac{d^{2\omega} p_1 d^{2\omega} p_2}{(2\pi)^{4\omega}}\frac{e^{i p_1 x+i p_2 y}}{p_1^2
p_2^2 (p_1+p_2)^2}\\
&=\frac{\Gamma(2\omega-3)}{64\pi^{2\omega}(\omega-1)}
\int_0^1
 d\alpha~~
 \frac{[\alpha(1-\alpha)]^{\omega-2}}{\left[\alpha (x-y)^2+(1-\alpha) y^2\right]^{2\omega-3}}\\
&\ \ \ \ \ \ \ \ \times{_2F_1}
\left(1,2\omega-3,\omega,\frac{({y}-\alpha{x})^2}{\alpha (x-y)^2+(1-\alpha) y^2}\right).
\end{split}
\end{equation}
From this representation, one obtains the behavior of $\mathcal{I}_1$ near $x=0$
\begin{equation}
\begin{split}
\mathcal{I}_1(0,y)=\frac{ \Gamma^2 (\omega -1)}{64 \pi ^{2 \omega }
 (2 \omega-3)(2-\omega)}
  \frac{1}{\left[
 {y}^2\right]^{2\omega-3}}.
 \end{split}
 \end{equation}
Since \eqref{inizio} is manifestly symmetric under the exchange
$x\leftrightarrow y$
and $x\leftrightarrow y-x$, the behavior at $y=0$ and $y=x$ is similar.
The integral  $\mathcal{I}_2(x,y)$ is defined as follows \cite{Bassetto:2008yf}
\begin{equation}
\label{pipp}
\begin{split}
 \mathcal{I}_2(x,y)
 &=-\frac{ \Gamma(2\omega-3) }{64\pi^{2\omega}(\omega-1)}
\!\!\!\int_0^1\!\!\!\!
 d\alpha \frac{\alpha^{\omega-1}(1-\alpha)^{\omega-2}}{\left[\alpha(1-\alpha){x}^2+
 ({y}-\alpha{x})^2\right]^{2\omega-3}}\\
&\ \ \ \ \ \ \ \ \ \ \ \ \ \ \ \ \ \ \ \ \ \ \ \ \ \ \ \ \ \times{_2F_1}
\left(1,2\omega-3,\omega,\frac{({y}-\alpha{x})^2}{({y}-\alpha{x})^2+
\alpha(1-\alpha){x}^2}\right).
 \end{split}
\end{equation}
Here we quote its  behavior  at $x=0$, $y=0$ and $y=x$.
\noindent

At ${x}=0$:
\begin{equation}
\begin{split}
 \mathcal{I}_2(0,y)
=-\frac{\Gamma^2(\omega-1)}{128\pi^{2\omega}(2-\omega)(2\omega-3)\left[
 ({y})^2\right]^{2\omega-3}}.
 \end{split}
\end{equation}
At $y=0$:
\begin{equation}
\begin{split}
\mathcal{I}_2(x,0)=-\frac{ \Gamma(2\omega-3) \Gamma(3-\omega)\Gamma(\omega-1)}{64\pi^{2\omega}\left[{x}\right]^{2\omega-3}}
\frac{ (\Gamma (\omega -2)-2  \Gamma (3-\omega ) \Gamma (2 \omega -4))}{4 (\omega -2)^3 \Gamma (2-\omega ) \Gamma (2 \omega
   -4)}. 
\end{split}
\end{equation}
At $y=x$:
\begin{equation}
\begin{split}
\mathcal{I}_2(x,x)
=-\frac{ \Gamma(2\omega-3)\Gamma(2-\omega)\Gamma(\omega) }{64\pi^{2\omega}(\omega-1)[{x}^2]^{2\omega-3}}
\frac{1-\frac{\Gamma (\omega -1)}{\Gamma (3-\omega ) \Gamma (2 \omega -2)}}{2 (\omega 
-2)}.
\end{split}
\end{equation}

In section \ref{sec:3} we introduced the following combination of the derivatives of
$\mathcal{I}_1$ and $\mathcal{I}_2$:
\begin{equation}
V^\mu(x,y)\equiv \frac{\partial \mathcal{I}_1({x},{y})}{\partial {x}_\mu}-
\frac{\partial \mathcal{I}_2({x},{y})}{\partial {y}_\mu}.
\end{equation}
Taking the derivative of \eqref{inizio} and \eqref{pipp}, $V^\mu$ can be 
expressed as \cite{Bassetto:2008yf}  
\begin{equation}
  \label{vumu}
\begin{split}
V^\mu(x,y)
&=-\frac{\Gamma(2\omega-2){x}^\mu}{32\pi^{{2\omega}}(\omega-1) ({x}^2)^{2\omega-2}}
\!\!\!\int_0^1\!\!\!\!
 d\alpha
{[\alpha(1-\alpha)]^{1-\omega}}{}
{_2F_1}(1,2\omega-2;\omega;\xi ) (1-\xi)^{2\omega-2},
\end{split}
\end{equation}
where
\begin{equation}
\xi=\frac{({y}-\alpha{x})^2}{({y}-\alpha{x})^2+
\alpha(1-\alpha){x}^2}.
\end{equation}
In particular, setting $\omega=2$ one has
\begin{equation}
  \begin{split}
    V^\mu (x,y)&=-\frac{{x}^\mu}{32\pi^4x^2}
\;\!\!\!\int_0^1\!\!\!\! d\alpha
\frac{1}{\alpha(1-\alpha){x}^2+(y-\alpha{x})^2}\\
 &=\frac{x^\mu}{32\pi^4 
 x^2}\frac{\log{\left[\frac{y^2}{(x-y)^2}\right]}}{(x-y)^2-y^2}  
 \end{split}
\end{equation}
For our purposes, however, it is more useful to rewrite $V^\mu$ as
\begin{equation}
\begin{split}\label{vuide}
 V^\mu(x,y)=&\frac{1}{32\pi^4 
 x^2}\biggr\{\frac{\partial}{\partial y_\mu}\biggr[\mathrm{Li}_2
 \left(1-\frac{(x-y)^2}{y^2}\right)+\frac 12 
 \log^2\left(\frac{(x-y)^2}{y^2}\right)\\
 &\qquad\qquad\qquad\qquad\qquad-\frac 12 
 \log^2\left(\frac{(x-y)^2}{x^2}\right)\biggr]-\frac{2(x-y)^\mu}{(x-y)^2}\log\left(\frac{y^2}{x^2}\right)\biggr\}.
\end{split}\end{equation}

\section{The latitude-north pole correlator}\label{sec:appB}

\begin{figure}[!h]
\centering
\includegraphics[width=4.6cm]{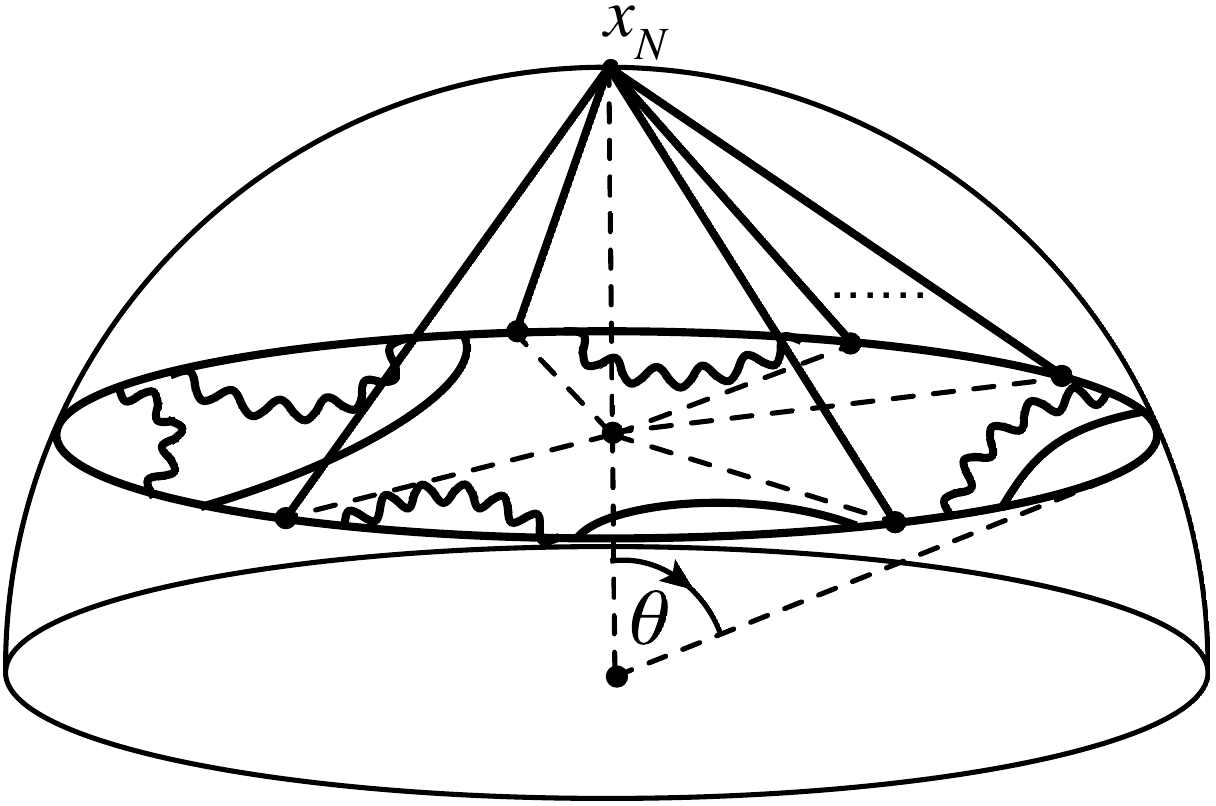}
\caption{A typical ladder diagram for latitude-north pole correlation function.}
\label{fig:ladderlat2}
\end{figure}
In this appendix we compute the ladder contribution to the correlation function 
of a Wilson loop lying on a latitude of $S^2$ and the chiral primary operator $\mathcal{O}_J$
inserted at the north pole to any order in $\lambda$.
As the generalized connection does not depend on the scalar field $\Phi^4$ (see 
\eqref{connection1}), the operator $\mathcal{O}_J$ effectively reduces to
\begin{equation}
\mathcal{O}_J(x_N)=\left(\frac{2\pi}{\sqrt{\lambda}}\right)^J\frac{1}{\sqrt{J}}\text{\text{Tr}}\left(\Phi_N^3\right)^J,
\end{equation}
where we use the notation $\Phi_N^3=\Phi^3(x_N)$.

We have
\begin{equation}\begin{split}\label{gpordinejlat}
\langle W[\mathcal{C}]\mathcal{O}_J(x_N)\rangle_{\text{ladder}}  
=\frac{1}{N}\sum_{n=0}^\infty\int_0^{2\pi}d\tau_1...\int_0^{\tau_{2n+J-1}}d\tau_{2n+J}
\,\langle 
\text{Tr}(\mathcal{A}_{1}...\mathcal{A}_{{2n+J}})\,\mathcal{O}_J(x_N)\rangle,
\end{split}\end{equation}
where $n$ counts
the number of ladder insertions in the Wilson loop.
Using the effective propagators \eqref{effprop1} we get
\begin{equation}\label{ntot}
\langle \text{Tr}(\mathcal{A}_{1}...\mathcal{A}_{{2n+J}})\,\text{Tr}(\Phi_N^3)^J\rangle=
\left(\frac{\lambda'}{16\pi^2}\right)^{n+J}\left(\frac{1}{1-\cos\theta}\right)^{J}\,N_{\text{tot}},
\end{equation}
where $N_{tot}$ is the total number of planar graphs.
Any such diagram originates from Wick contractions of this kind
\small\begin{equation}\label{schema}
\bcontraction[2ex]{}{\mathcal{A}_{_1}}{\overbrace{\mathcal{A}_{_2}...\mathcal{A}_{_{2s_1+1}}}^{2s_1}\mathcal{A}_{_{2s_1+2}}
\overbrace{\mathcal{A}_{_{2s_1+3}}...\mathcal{A}_{_{2s_1+2s_2+2}}}^{2s_2}...{//}...
\overbrace{\mathcal{A}_{_{2n+J-2s_{J-1}}}...\mathcal{A}_{_{2n+J-1}}}^{2s_{J-1}}\mathcal{A}_{_{2n+J}}
\qquad}{\Phi^3_{_N}}
\bcontraction[3ex]{\mathcal{A}_{_1}\overbrace{\mathcal{A}_{_2}...\mathcal{A}_{_{2s_1+1}}}^{2s_1}}
{\mathcal{A}_{_{2s_1+2}}}{\overbrace{\mathcal{A}_{_{2s_1+3}}...\mathcal{A}_{_{2s_1+2s_2+2}}}^{2s_2}...{//}...
\overbrace{\mathcal{A}_{_{2n+J-2s_{J-1}}}...\mathcal{A}_{_{2n+J-1}}}^{2s_{J-1}}\mathcal{A}_{_{2n+J}}
\qquad\Phi^3_{_N}}{\Phi^3_{_N}}
\bcontraction[4ex]{\mathcal{A}_{_1}\overbrace{\mathcal{A}_{_2}...\mathcal{A}_{_{2s_1+1}}}^{2s_1}\mathcal{A}_{_{2s_1+2}}
\overbrace{\mathcal{A}_{_{2s_1+3}}...\mathcal{A}_{_{2s_1+2s_2+2}}}^{2s_2}...{//}...
\overbrace{\mathcal{A}_{_{2n+J-2s_{J-1}}}...\mathcal{A}_{_{2n+J-1}}}^{2s_{J-1}}}
{\mathcal{A}_{_{2n+J}}}{\qquad\Phi^3_{_N}\Phi^3_{_N}...{//}...}{\Phi^3_{_N}}
\mathcal{A}_{_1}\overbrace{\mathcal{A}_{_2}...\mathcal{A}_{_{2s_1+1}}}^{2s_1}\mathcal{A}_{_{2s_1+2}}
\overbrace{\mathcal{A}_{_{2s_1+3}}...\mathcal{A}_{_{2s_1+2s_2+2}}}^{2s_2}...{//}...
\overbrace{\mathcal{A}_{_{2n+J-2s_{J-1}}}...\mathcal{A}_{_{2n+J-1}}}^{2s_{J-1}}\mathcal{A}_{_{2n+J}}
\qquad\overbrace{\Phi^3_{_N}\Phi^3_{_N}...{//}...\Phi^3_{_N}}^{J}
\end{equation}
\normalsize
where $\sum_{i=1}^{J-1}s_i=n$ and the $2s$ generalized connections $\mathcal{A}_i$
under the brackets have to be contracted among themselves.
According to \cite{Erickson:2000af} the number of these planar contractions is
\begin{equation}
N_s=\frac{(2s)!}{(s+1)!s!}.
\end{equation}
Therefore the total number of planar graphs is
\begin{equation}\label{babauk}
  \begin{split}
N_{\text{tot}}=(2n+J)\sum_{s_1=0}^n N_{s_1}\sum_{s_2=0}^{n-s_1}N_{s_2}\sum_{s_3=0}^{n-s_1-s_2}...//
...\sum_{s_{J-1}=0}^{n-\sum_{i=1}^{J-2}s_i}N_{s_{J-1}}N_{n-\sum_{i=1}^{J-2}s_i-s_{J-1}},
\end{split}\end{equation}
where the factor $(2n+J)$ comes from the ciclicity of the trace.
Using the recurrence relation~\cite{Erickson:2000af}
\begin{equation}\label{ricorsiva}
N_{n+1}=\sum_{k=0}^nN_{n-k}\,N_k,
\end{equation}
with $N_0=1$, 
we can perform the sum over $s_{J-1}$, obtaining
\small\begin{equation}\label{numerografk}
  \begin{split}
&(2n+J)\sum_{s_1=0}^n\frac{2s_1!}{s_1!(s_1+1)!}\sum_{s_2=0}^{n-s_1}\frac{2s_2!}{s_2!(s_2+1)!}
\sum_{s_3=0}^{n-s_1-s_2}...//...\\
&\times\left[\sum_{s_{J-2}=0}^{n-\sum_{i=1}^{J-3}s_i+1}N_{s_{J-2}}N_{n-\sum_{i=1}^{J-3}s_i+1-s_{J-2}}-
\frac{(2n-2\sum_{i=1}^{J-3}s_i+2)!}{(n-\sum_{i=1}^{J-3}s_i+1)!(n-\sum_{i=1}^{J-3}s_i+2)!}\right].
\end{split}
\end{equation}\normalsize
Iterating this process $(J-3)$-times, we get
\begin{equation}
N_{\text{tot}}=\frac{J(2n+J)!}{n!(n+J)!}.
\end{equation}
Substituting this result into the \eqref{ntot} and performing the trivial
integrations, we can write
\begin{equation}\begin{split}\label{laddertutto}
\langle{W[\mathcal{C}]\,\mathcal{O}_J(x_N)}\rangle_{\text{ladder}}=&\frac{1}{N}\frac{\sqrt{J}}{2^J}\left(\frac{A_2}{A_1}\right)^{J/2}
\sum_{n=0}^\infty\frac{1}{n!(n+J)!}\left(\frac{\sqrt{\lambda'}}{2}\right)^{2n+J}\\
=&\frac{1}{N}\frac{\sqrt{J}}{2^J}\left(\frac{A_2}{A_1}\right)^{J/2}
I_J(\sqrt{\lambda'}).
\end{split}\end{equation}
Thus the sum of all ladder contribution reproduces the localization result \eqref{matrmodel}.

\section{Summing up interactions I: the details}\label{sec:appC}

\subsection*{The evaluation of $P_1$}
Using the integral representation of $\mathcal{I}_1$ given in \eqref{inizio}, we get
\small
\begin{equation}\begin{split}
P_1=&\oint d\tau_1\,d\tau_2
\;\;\mathcal{I}_1(x_1-x_2,x_N-x_2)\;(x_1-x_2)^2\\
=&\frac{\Gamma(2\omega-3)}{2^5\pi^{2\omega}(\omega-1)}\frac{\sin^2\theta}{(1-\cos\theta)^{2\omega-3}}\oint
d\tau_1 d\tau_2\int_0^1d\alpha
\;\frac{[\alpha(1-\alpha)]^{\omega-2}(1-\cos\tau_{12})}{2^{2\omega-3}}\\
&\qquad\qquad\qquad\qquad\qquad\qquad\qquad\times \;{_2F_1}
\biggr(1,2\omega-3;\omega;1-\alpha(1-\alpha)\frac{\sin^2\theta}{(1-\cos\theta)}(1-\cos\tau_{12})\biggr).
\end{split}\end{equation}\normalsize
With the help of the following identity
\begin{equation}
  \begin{split}
{_2F_1}(\alpha,\beta;\gamma;z)=&\frac{\Gamma(\gamma)\Gamma(\gamma-\alpha-\beta)}{\Gamma(\gamma-\alpha)\Gamma(\gamma-\beta)}
\;{_2F_1}(\alpha,\beta;\alpha+\beta-\gamma+1;1-z)\\
&+(1-z)^{\gamma-\alpha-\beta}\frac{\Gamma(\gamma)\Gamma(\alpha+\beta-\gamma)}{\Gamma(\alpha)\Gamma(\beta)}
\;{_2F_1}(\gamma-\alpha,\gamma-\beta;\gamma-\alpha-\beta+1;1-z),
  \end{split}
\end{equation}
and the series representation of the hypergeometric function, $P_1$ becomes
\small
\begin{equation}
  \begin{split}
\frac{1}{2^{6+2\epsilon}\pi^{4+2\epsilon}}
\sum_{k=0}^\infty \oint
& d\tau_1 d\tau_2\int_0^1d\alpha
\biggr[\frac{(\sin^2\theta)^{1-\epsilon+k}}{(1-\cos\theta)^{1+\epsilon+k}}
 \Gamma(\epsilon)\frac{\Gamma(1+\epsilon+k)}{\Gamma(k+1)}[\alpha(1-\alpha)]^{k}(1-\cos\tau_{12})^{1-\epsilon+k}\\
&+\frac{(\sin^2\theta)^{1+k}}{(1-\cos\theta)^{1+2\epsilon+k}}
\;\frac{\Gamma(-\epsilon)}{\Gamma(1-\epsilon)}\frac{\Gamma(1+2\epsilon+k)\Gamma(1+\epsilon)}{\Gamma(1+\epsilon+k)}[\alpha(1-\alpha)]^{\epsilon+k}(1-\cos\tau_{12})^{k+1}\biggr].
\end{split}
\end{equation}
\normalsize
The integrations can now be performed easily, and we obtain
\small
\begin{equation}
  \begin{split}
\frac{\sqrt{\pi}}{2^{6+2\epsilon}\pi^{3+2\epsilon}}
\sum_{k=0}^\infty&\biggr[2^{3+k}\frac{(\sin^2\theta)^{1+k}}{(1-\cos\theta)^{1+2\epsilon+k}}\frac{\Gamma(-\epsilon)}{\Gamma(1-\epsilon)}\frac{\Gamma(1+2\epsilon+k)\Gamma(1+\epsilon)\Gamma(1+\epsilon+k)\Gamma(3/2+k)}{\Gamma(2+2\epsilon+2k)\Gamma(k+2)}\\
 &\;+2^{3-\epsilon+k}\frac{(\sin^2\theta)^{1-\epsilon+k}}{(1-\cos\theta)^{1+\epsilon+k}}\Gamma(\epsilon)\frac{\Gamma(1+\epsilon+k)\Gamma(k+1)\Gamma(3/2-\epsilon+k)}{\Gamma(2k+2)\Gamma(2-\epsilon+k)}\biggr].
  \end{split}
\end{equation}
Taking the limit $\epsilon\rightarrow 0$
we see that divergences cancel, and the sum over $k$  gives
\begin{equation}\begin{split}
P_1=\frac{1}{8 \pi ^2}\Biggr[\frac{\pi^2}{6}-\mathrm{Li}_2\left(\sin^2\frac\theta 
2\right)\Biggr].
\end{split}\end{equation}

\subsection*{The evaluation of $P_2$}
Using the integral representation of $\mathcal{I}_1$ and $\mathcal{I}_2$ given in \eqref{inizio} and \eqref{pipp}, we get
\begin{equation}
  \begin{split}
P_2=&\oint d\tau_1\,d\tau_2
\;\biggr[\mathcal{I}_1(x_1-x_N,x_2-x_N)+\mathcal{I}_2(x_2-x_N,x_1-x_N)\biggr]\\
=&\frac{1}{2^{2\omega+3}}
\frac{\Gamma(2\omega-3)}{\pi^{2\omega}(\omega-1)}
\frac{1}{(1-\cos\theta)^{2\omega-3}}
\oint d\tau_1 d\tau_2
\int_0^1
 d\alpha~~
 \frac{(1-\alpha)[\alpha(1-\alpha)]^{\omega-2}}{\left[1+\alpha\cos\theta-\alpha(1+\cos\theta)\cos\tau_{21}\right]^{2\omega-3}}\\
&\qquad \qquad \qquad \qquad \ \ \ \ \ \times{_2F_1}
\left(1,2\omega-3,\omega,1-\frac{\alpha(1-\alpha)}{1+\alpha\cos\theta-\alpha(1+\cos\theta)\cos\tau_{21}}\right).
  \end{split}
\end{equation}
With the help of the identity
\begin{equation}
_2F_1(\alpha,\beta;\gamma;z)=(1-z)^{-\beta}\,_2F_1\left(\beta,\gamma-\alpha;\gamma;\frac{z}{z-1}\right),
\end{equation}
we arrive to the following expression
\begin{equation}
  \begin{split}\label{premb}
    P_2=&\frac{1}{2^{2\omega+3}}
\frac{\Gamma(2\omega-3)}{\pi^{2\omega}(\omega-1)}
\frac{1}{(1-\cos\theta)^{2\omega-3}}
\oint d\tau_1 d\tau_2
\int_0^1
 d\alpha~~
 \alpha^{1-\omega}(1-\alpha)^{2-\omega}\\
&\qquad \qquad \qquad \ \ \ \ \ \times{_2F_1}
\left(2\omega-3,\omega-1,\omega,-\frac{1-\alpha}{\alpha}-\frac{(1+\cos\theta)(1-\cos\tau_{21})}{1-\alpha}\right).
  \end{split}
\end{equation}
Exploiting the Mellin-Barnes representation of the hypergeometric
function, we recast $P_2$ as
\begin{equation}
  \begin{split}\label{MB1}
    &\frac{1}{2^{2\omega+4}}
\frac{1}{\pi^{2\omega+1}\,i}
\frac{1}{(1-\cos\theta)^{2\omega-3}}
\oint d\tau_1 d\tau_2
\int_0^1
 d\alpha~~
 \alpha^{1-\omega}(1-\alpha)^{2-\omega}\\
&\qquad \qquad \times\int_{-i\infty}^{i\infty} dt
\;\frac{\Gamma(2\omega-3+t)\Gamma(\omega-1+t)\Gamma(-t)}{\Gamma(\omega+t)}
\left(\frac{1-\alpha}{\alpha}+\frac{(1+\cos\theta)(1-\cos\tau_{21})}{1-\alpha}\right)^t.
\end{split}
\end{equation}
We perform a Mellin-Barnes transform also of last factor in \eqref{MB1}, 
obtaining
\begin{equation}
  \begin{split}
    -&\frac{1}{2^{2\omega+5}}
\frac{1}{\pi^{2\omega+2}}
\frac{1}{(1-\cos\theta)^{2\omega-3}}
\int_{-i\infty}^{i\infty} dt\int_{-i\infty}^{i\infty}ds
\;\frac{\Gamma(2\omega-3+t)\Gamma(s)\Gamma(-t-s)}{(\omega-1+t)}(1+\cos\theta)^{-s}\\
&\qquad\qquad\qquad\qquad\times\oint d\tau_1\,d\tau_2\;(1-\cos\tau_{21})^{-s}
\int_0^1 d\alpha~~
\,\alpha^{1-\omega-t-s}(1-\alpha)^{2-\omega+t+2s}.
\end{split}
\end{equation}
Evaluating the integrals over $\tau_1$, $\tau_2$ and $\alpha$ and setting 
$\omega=2+\epsilon$, we get
\small
\begin{equation}
\label{hahahah}
  \begin{split}
-\frac{1}{2^{7+2\epsilon}}
\frac{\sqrt{\pi}}{\pi^{5+2\epsilon}}
\int_{-i\infty}^{i\infty} dt\int_{-i\infty}^{i\infty} ds\frac{(1+\cos\theta)^{-s}}{(1-\cos\theta)^{1+2\epsilon}}
\;2^{-s}&\frac{\Gamma(1+2\epsilon+t)\Gamma(s)\Gamma(-t-s)\Gamma(\frac{1}{2}-s)}{(1+\epsilon+t)\Gamma(1-s)\Gamma(s-2\epsilon+1)}\\
&\qquad\times\Gamma(-\epsilon-t-s)\Gamma(1+t+2s-\epsilon).
\end{split}
\end{equation}\normalsize
Analyzing the singularity structure of the integrand, it is possible 
to choose a contour of integration for $s$ and $t$ that allows us to take $\epsilon=0$
and satisfies
\begin{equation}
0<Re(s)<1/2,\qquad\quad -1<Re(t)<-Re(s).
\end{equation}
Shifting $t\rightarrow t-s$ and expressing everything in terms of
$\Gamma$-functions (\ref{hahahah}) becomes
\small\begin{equation}\begin{split}
-\frac{1}{2^{7}}\frac{\sqrt{\pi}}{\pi^{5}}
\int_{-i\infty}^{i\infty}ds&
\;\frac{(1+\cos\theta)^{-s}}{(1-\cos\theta)}\frac{2^{-s} \sin (\pi  s) \Gamma \left(\frac{1}{2}-s\right) \Gamma (s)}{\pi  
s}\\
&\times\int_{-i\infty}^{i\infty} dt\; \biggr[\Gamma (t-s) \Gamma (-t)^2 \Gamma (s+t+1)-
\frac{\Gamma (t-s) \Gamma (-t)^2 \Gamma 
(s+t+1)\Gamma(t-s+1)}{\Gamma(t-s+2)}\biggr].
\end{split}\end{equation}\normalsize
The integral over $t$ can be performed using the two Barnes lemmas, obtaining
\begin{equation}\label{ultimo}
\frac{i}{2^{6}}
\frac{\sqrt{\pi}}{\pi^{4}}
\int_{-i\infty}^{i\infty}ds
\;\frac{(1+\cos\theta)^{-s}}{(1-\cos\theta)}
2^{-s}\Gamma \left(\frac{1}{2}-s\right) \Gamma(s)^2\Gamma(-s)
\left(1-\frac{\Gamma(1-s)^2}{\Gamma(1-2s)}\right).
\end{equation}
Finally the integral over $s$ is easily done through residue theorem, and we get
\small
\begin{equation}\begin{split}
P_2=\frac{1}{96\pi^2}\frac{1}{1-\cos\theta}\biggr[\pi^2-3\,\mathrm{Li}_2\left(\sin^2\frac\theta2\right)-6\left(\log^2\left(\cos\frac\theta 
2\right)+\arctan^2\left(\sqrt{1+2\cos\theta}\right)\right)\biggr].
\end{split}\end{equation}
\normalsize

\subsection*{The evaluation of $P_3$}
The evaluation of $P_3$ is straightforward: taking $\omega=2+\epsilon$ we get
\begin{equation}\begin{split}
P_3=&\oint d\tau_1\,d\tau_2
\;\biggr[\mathcal{I}_2(x_2-x_N,x_2-x_N)+\mathcal{I}_1(0,x_2-x_N)\biggr]\\
=&-\frac{\csc (\pi  \epsilon ) (\Gamma (\epsilon )-2 \Gamma (1-\epsilon ) \Gamma (2 \epsilon ))}{2^{6+2\epsilon}\pi^{1+2\epsilon}(1-\cos\theta)^{1+2\epsilon}\Gamma (1-\epsilon )}\\
=& \frac{1}{192}\frac{1}{1-\cos\theta}\qquad\epsilon\rightarrow 0.
\end{split}
\end{equation}

\subsection*{The evaluation of $P_4$}
We have
\begin{equation}\begin{split}
P_4=&\oint d\tau_1 d
\tau_3\;\mathrm{Li}_2\left(1-\frac{\sin^2\theta}{1-\cos\theta}(1-\cos\tau_{31})\right)\\
=&8\pi\int_0^{\pi/2} d\tau
\;\mathrm{Li}_2\left(1-K(\theta)\sin^2\tau\right),
\end{split}\end{equation}
where $K(\theta)=4\cos^2\frac\theta 2$. Using the integral representation
of the dilogarithm and changing variable to $x=\sin\tau$, we obtain
\small
\begin{equation}\begin{split}
P_4=&-8\pi\int_0^{1}ds
\int_0^{1} dx
\frac{\log\left[1-s(1-K(\theta)x^2)\right]}{s\sqrt{1-x^2}}\\
=&\frac{2}{3}\pi^4-8\pi^2\int_0^{1}
\frac{ds}{s}\log\biggr[\frac12\left(1+\sqrt{1+\frac{K(\theta)s}{1-s}}\right)\biggr]\\
=&\frac{\pi^4}{6}-2\pi^2\log^2(2)-8\pi^2\int_2^K dK'\;\frac{d}{dK'}\int_0^{1}
\frac{ds}{s}\log\biggr[\frac12\left(1+\sqrt{1+\frac{K'[\theta]s}{1-s}}\right)\biggr].
\end{split}\end{equation}
\normalsize
Taking the derivative and interchanging the order of integration we get
\small\begin{equation}\begin{split}
P_4=\frac23 \pi^4-8\pi^2\left[\log^2{\left(\cos\frac\theta 
2\right)}+\arctan^2{\left(\sqrt{1+2\cos\theta}\right)}\right].
\end{split}\end{equation}\normalsize

\section{Some useful integrals}\label{sec:appD}
In this appendix we give the integrals $\mathcal{J}_n$ and $\tilde{\mathcal{J}}$ 
needed to evaluate \eqref{o2xop3}:
\begin{equation}
\mathcal{J}_n=\int_0^{2\pi}d\tau\,\tau^n\,f(\tau),\qquad\tilde{\mathcal{J}}=\int_0^{2\pi}d\tau_1\,\tau_1\;f(\tau_1)\int_{\tau_1}^{2\pi}d\tau_2\,f(\tau_2),
\end{equation}
with $f(\tau)$ given in \eqref{ftau} (henceforth we set $\sin\phi=b$, thus 
$\sigma=\sqrt{\frac{1+b}{1-b}}$ ).

The integral $\mathcal{J}_0$ and $\mathcal{J}_1$ are straightforward. Making the 
change of variables $\tan{\frac\tau 2}=x$, $\mathcal{J}_0\;$�becomes
\begin{equation}
\mathcal{J}_0=\frac{2}{1-b}\int_0^{\infty} \frac{dx}{1+\sigma^2x^2}+(b\rightarrow 
-b)=\frac{2\pi}{\sqrt{1-b^2}}.
\end{equation}
For $\mathcal{J}_1$, periodicity of $f(\tau)$ allows us to write
\begin{equation}
\label{i1}
\mathcal{J}_1=\pi\int_{-\pi}^{\pi} d\tau\; 
\frac{1}{1+b\cos{\tau}}=\pi\mathcal{J}_0.
\end{equation}
The evaluation of $\mathcal{J}_2$ is a bit tricky. Again using periodicity and 
the change of variables $\tan{\frac\tau 2}=x$, we can write 
\begin{equation}\label{fff}
\mathcal{J}_2=\pi^2\mathcal{J}_0+\frac{16}{\sqrt{1-b^2}}\mathcal{F}(\sigma),
\end{equation}
where
\begin{equation}
\mathcal{F}(\sigma)=\int_0^\infty dx\frac{\arctan^2{\sigma x}}{1+x^2}.
\end{equation}
First we evaluate the derivative of $\mathcal{F}(\sigma)$
\begin{equation}
  \begin{split}
    \mathcal{F}'(\sigma)=\frac{1}{\sigma^2}\int_{-\infty}^{\infty} dx \frac{x\arctan{\sigma 
    x}}{(1+x^2)(1/\sigma^2+x^2)}=\frac{\pi}{\sigma^2-1}\log{\left(\frac{1+\sigma}{2}\right)},
  \end{split}
\end{equation}
and then we write
\begin{equation}\label{f2sigma}
  \begin{split}
    \mathcal{F}(\sigma)=&\mathcal{F}(1)+\int_1^\sigma d\sigma' \mathcal{F}'(\sigma')\\
=&\frac13\left(\frac{\pi}{2}\right)^3-\frac{\pi}{2}\left[ 
\frac12\log^2{\left(\frac{1+\sigma}{2}\right)}+\text{Li}_2\left(\frac{1-\sigma}{2}\right)\right].
  \end{split}
\end{equation}
Finally, substituting this result in (\ref{fff}), we obtain:
\begin{equation}\label{i2}
\mathcal{J}_2=\frac{4\pi}{\sqrt{1-b^2}}\left[\frac23 
\pi^2-2\text{Li}_2\left(\frac{1-\sigma}{2}\right)-
\log^2\left(\frac{1+\sigma}{2}\right)\right].
\end{equation}  

The integral $\tilde{\mathcal{J}}$ can be treated in a similar way, with the 
change of variables $x_{1,2}=\cot\left(\frac{\tau_{1,2}}{2}\right)$, one has
\begin{equation}
  \begin{split}
\tilde{\mathcal{J}}=\frac{8}{(1+b)^2}\int_{-\infty}^{\infty}dx_1\,\frac{\text{arccot}(x_1)}{1+\frac{x_1^2}{\sigma^2}}
\int_{-\infty}^{x_1}dx_2\frac{1}{1+\frac{x_2^2}{\sigma^2}}.
\end{split}
\end{equation}
Performing the integration over $x_2$ and integrating by parts we get:
\begin{equation}
  \begin{split}
\tilde{\mathcal{J}}=&\frac{\pi^3}{1-b^2}+\frac{8}{1-b^2}
\int_{-\infty}^{\infty}dx_1\,\frac{\arctan(x_1/\sigma)^2}{1+x_1^2}\\
=&\frac{\pi^3}{1-b^2}-\frac{8}{1-b^2}\mathcal{F}\left(\frac 1\sigma\right),
  \end{split}
\end{equation}
where $\mathcal{F}\left(\sigma\right)$ is given in (\ref{f2sigma}).

\section{Summing up interactions II: the details}\label{sec:appE}

\begin{figure}[!ht]
    \centering
    \includegraphics[width=11cm]{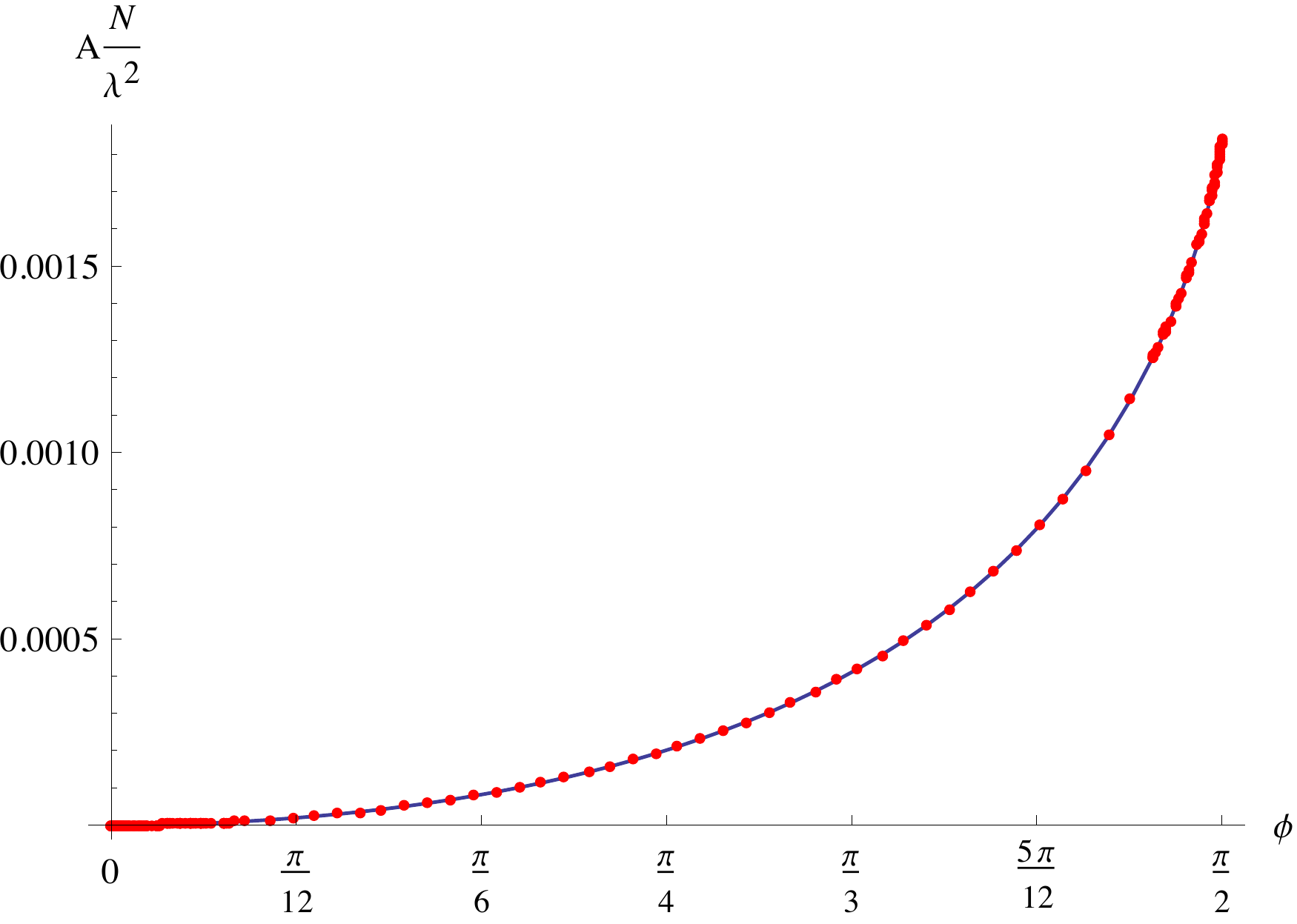}
    \caption{Numerical evaluation of the quantity $\text{A}\frac{N}{\lambda^2}$ as a function of the angle $\phi$ 
    obtained with Wolfram Mathematica routine NIntegrate.}\label{fig:intnum}
\end{figure}

In this appendix we evaluate the contribution to (\ref{somma2}), called A, i.e.
\begin{equation}
  \begin{split}\label{e1}
\text{A}=&-\frac{\lambda^2\cos^2\phi}{2^3\sqrt{2}N}\oint
d\tau_1 d\tau_2\,f(\tau_1)\,f(\tau_2)
\;\;\mathcal{I}_1(x_1-x_{2},x_{\mathcal{O}}-x_2)\;(x_1-x_2)^2\\
&+\frac{\lambda^2\cos^2\phi}{2\sqrt{2}N}\oint d\tau_1 d\tau_2\,f(\tau_1)
\;\biggr[\mathcal{I}_1(x_1-x_{\mathcal{O}},x_2-x_{\mathcal{O}})+\mathcal{I}_2(x_2-x_{\mathcal{O}},x_1-x_{\mathcal{O}})\biggr]\\
&-\frac{\lambda^2\cos^2\phi}{2^8\pi^4\sqrt{2}N}\oint d\tau_1 
d\tau_2\,f(\tau_1)\,f(\tau_2)
\Biggr[\mathrm{Li}_2\biggr(1-(1-\cos\tau_{21})f(\tau_1)\biggr)\Biggr],
\end{split}
\end{equation}
with $\mathcal{I}_1$ and $\mathcal{I}_2$ given in\eqref{inizio} and 
\eqref{pipp}.
The new feature of this contribution with respect to the integrals evaluated in 
appendix \ref{sec:appC} is the appearance of the functions $f(\tau_i)$ (also in the argument of
the dilogarithm and the hypergeometric function). Because of this fact, 
we are not able to compute \eqref{e1} analytically and we have to resort to its 
numerical evaluation for different values of the angle $\phi\in [0,\pi/2]$, 
which identifies the position of the operator on the sphere. 
The results are shown in Figure \ref{fig:intnum}.

In particular, the vanishing of A at $\phi=0$ (i.e. the operator on the north-pole) 
is consistent with analytic results of section \ref{sec:3}.
 
The last integral in (\ref{somma2}), i.e. the term B, is
\begin{equation}
B=\frac{\lambda^2\cos^2\phi}{2^9\pi^4\sqrt{2}N}\oint d\tau_1 d\tau_2 d
\tau_3\,\epsilon(\tau_1,\tau_2,\tau_3)\,f(\tau_1)\,f(\tau_3) 
\,F(\tau_3,\tau_2),
\end{equation}
with
\begin{equation}
F(\tau_3,\tau_2)=F(\tau_2,\tau_3)=\cot\left(\frac{\tau_{32}}{2}\right)\log\left(\frac{f(\tau_3)}{f(\tau_2)}\right).
\end{equation}
It is useful to express $f(\tau)$ in terms of its primitive $g(\tau)$
\begin{equation}
g(\tau)=\frac{2}{\sqrt{1-b^2}}\text{arccot}{\left(\frac1\sigma\cot{\frac{\tau_i}{2}}\right)},
\end{equation}
with
\begin{equation}\label{g2pi}
g(0)=\lim_{\tau \to 0^+}g(\tau)=0,\qquad\qquad g(2\pi)=\lim_{\tau\rightarrow 
2\pi^-}g(\tau)=\frac{2\pi}{\sqrt{1-b^2}}.
\end{equation}
Then using the integration by parts and \eqref{epsilon} and \eqref{eta},
we can evaluate one of the three integrals, obtaining
\begin{equation}
  \begin{split}\label{a10}  
B=\frac{\lambda^2\cos^2\phi}{2^8\pi^4\sqrt{2}N}\oint d\tau_2\biggr\{&\oint d\tau_3
\,(g(\tau_3)-g(\tau_2))\,f(\tau_3)\,F(\tau_3,\tau_2)\\
+&\frac12g(2\pi)\biggr[\int_0^{\tau_2}d\tau_3    
\,f(\tau_3)\,F(\tau_3,\tau_2)-\int_{\tau_2}^{2\pi}d\tau_3 
\,f(\tau_3)\,F(\tau_3,\tau_2)\biggr]\biggr\}.
  \end{split}
\end{equation}
With the usual change of variables $x=\cot{\frac{\tau}{2}}$, in both integrals, 
we can evaluate one of the two integrals, obtaining
\begin{equation}\label{Bint}
B=\frac{\lambda^2}{2^4\pi^2\sqrt{2}N}\biggr\{\log^2\left(\frac{2\sigma}{\sigma+1}\right)-
\frac{1}{2\pi}\int_0^{\infty}dy\,\frac{\sigma\,\log^2\left(\frac{1+y^2}{\sigma^2+y^2}\right)}{\sigma^2+y^2}\biggr\}.
\end{equation}
The integral in \eqref{Bint} is done by expanding the integrand in power series
in $\sigma$ at $\sigma=1$
\begin{equation}
  \begin{split}\label{series1}
\int_0^{\infty}dy\,\frac{\sigma\,\log^2\left(\frac{1+y^2}{\sigma^2+y^2}\right)}{\sigma^2+y^2}=
\int_{-\infty}^\infty&\,dy\;\sum_{n=2}^{\infty}\sum_{j=1}^{n-1}
\frac{i^{n+1}}{2(n-j)\left(y^2+1\right)^{n+1}}((i-y)^{n-j}+(i+y)^{n-j})\\
&\times(i-y)^{j+1} \left(H_j-\beta_{\frac{i+y}{i-y}}(1+j,0)-
\log \left(\frac{2 y}{y-i}\right)\right) (\sigma -1)^n,
  \end{split}
\end{equation} 
where $H_n=\sum_{k=1}^n\frac{1}{k}$ are the harmonic numbers and 
$\beta_z(a,b)=z^a\sum_{n=0}^\infty\frac{(1-b)_n}{n!(a+n)}z^n$ is the incomplete $\beta$-function.

Given the following series expansion
\begin{equation}
\left(H_j-\beta_{\frac{i+y}{i-y}}(1+j,0)-
\log \left(\frac{2 
y}{y-i}\right)\right)=\sum_{k=1}^j\frac{1}{k}\left(\frac{(i+y)^k}{(i-y)^k}+1\right),
\end{equation}
\eqref{series1} becomes 
\begin{equation}
  \begin{split}\label{sumsumsum}
&\sum_{n=2}^{\infty}\sum_{j=1}^{n-1}\sum_{k=1}^j
\int_{-\infty}^\infty\,dy\;\frac{i^{n+1}}{2(n-j)\left(y^2+1\right)^{n+1}}((i-y)^{n-j}+(i+y)^{n-j})\frac{1}{k}\left(\frac{(i+y)^k}{(i-y)^k}+1\right)(\sigma 
-1)^n\\
=&\sum_{n=2}^{\infty}\sum_{j=1}^{n-1}\sum_{k=1}^j
 \frac{(\sigma -1)^n}{\pi ^2 k \Gamma (k+1) (n-j) \Gamma (k+n-j)}\biggr[i e^{-i \pi  k} 2^{-n-2} \sin (\pi  j) \sin (\pi  (k-j))\\
 &\qquad\times\biggr(\pi ^2 (-k) \csc (\pi  j) \csc (\pi  (k-j)) \Gamma (k+n-j) \left(\pi  e^{i \pi  k} (-2)^n \csc (\pi  k) \, _2\tilde{F}_1\left(1,1-n;2-k;\frac{1}{2}\right)\right.\\
 &\qquad\left.+2 \left(-1+e^{2 i \pi  k}\right) \Gamma (n) \Gamma (k-n)\right)-\pi  e^{i \pi  k} \Gamma (k+1) \sin (\pi  j) \left(\pi  \csc ^2(\pi  j) \csc (\pi  (k-j))\right.\\
 &\qquad\left.\times \left(\pi  2^n \csc (\pi  (k-j)) \, _2\tilde{F}_1\left(1,1-n;-k-n+j+2;\frac{1}{2}\right)+2 (-1)^n \Gamma (k+n-j)\right.\right.\\
 &\qquad\left.\left.\times\left(\beta_{\frac{1}{2}}(k-j,-k-n+j+1)+\beta_{\frac{1}{2}}(k-n,1-k)+\beta_{\frac{1}{2}}(-j,-n+j+1)+\beta_{\frac{1}{2}}(-n+j+1,-j)\right)\right)\right.\\
 &\qquad\left.+4 i \Gamma (n) \left(\pi  (-1)^n \csc ^2(\pi  j) \Gamma (k-j)+\Gamma (-j) \csc (\pi  (k-j)) \Gamma (-n+j+1) \Gamma 
 (k+n-j)\right)\right)\biggr)\biggr]\\
 =&\sum_{n=2}^\infty \frac{2^{-n} \pi e^{i \pi  n}}{n^2}\biggr[2^n n \left(n \, _3F_2\left(1,1,1-n;2,2;\frac{1}{2}\right)+2 (\Phi (2,1,n)+\psi ^{(0)}(n)+\gamma )-\log (4)\right)\\
 &\qquad\qquad\qquad\qquad+2 n \Phi \left(\frac{1}{2},1,n\right)+2 i \pi  
 n-2\biggr] (\sigma -1)^n\\
 =&-\pi\biggr[2 \text{Li}_2\left(\frac{1-\sigma }{2}\right)+2 \text{Li}_2\left(\frac{\sigma -1}{2 \sigma }\right)-\log (\sigma ) (\log (\sigma )-2 \log (\sigma +1)+\log 
 (4))\biggr],
 \end{split}
\end{equation}
where $\gamma$ is the Eulero-Mascheroni constant, $_2\tilde{F}_1(a,b,c,z)=\frac{_2F_1(a,b,c,z)}{\Gamma(c)}$ is the 
regularized hypergeometric function,
$\Phi(z,a,s)=\sum_{n=0}^\infty \frac{z^n}{(a+n)^s}$ is the Lerch trascendent 
function
and $\psi ^{(0)}(z)=\frac{d}{dz}\log \Gamma(z)$ is the digamma function.

Finally, including the result \eqref{sumsumsum} in \eqref{Bint}, we obtain
\small\begin{equation}\label{Bfin}
  \begin{split}
B=&\frac{\lambda^2}{2^4\pi^2\sqrt{2}N}\biggr\{\log^2\left(\frac{2\sigma}{\sigma+1}\right)+
\mathrm{Li}_2\left(\frac{1-\sigma}{2}\right)+\mathrm{Li}_2\left(\frac{\sigma-1}{2\sigma}\right)
-\frac12\log (\sigma )\log \left(\frac{4\sigma}{(\sigma+1)^2} \right)\biggr\}\\
=&\frac{\lambda^2}{2^5\pi^2 \sqrt{2} N}\biggr[\log^2\left(\frac{2\sigma}{1+\sigma}\right)+
\log^2\left(\frac{1+\sigma}{2}\right)+2\text{Li}_2\left(\frac{1-\sigma}{2}\right)+2\text{Li}_2\left(\frac{\sigma-1}{2\sigma}\right)\biggr]\\
=&-2\text{\textbf{L}}.
\end{split}
\end{equation}
\normalsize
where \textbf{L} is defined in \eqref{L}.

\phantomsection

\end{document}